\documentclass[prd,preprint,superscriptaddress,
  tightenlines,nofootinbib, eqsecnum]{revtex4-2}

\usepackage{amsmath}
\usepackage{amsfonts}
\usepackage{amssymb}
\usepackage{bm}
\usepackage{hyperref}
\usepackage{mathrsfs}
\usepackage{graphicx}

\usepackage{empheq}

\usepackage{ulem}
\usepackage[usenames]{color}

\definecolor{darkgreen}{rgb}{0,0.5,0}

\hypersetup{
 bookmarks=true,         
 unicode=false,          
 pdftoolbar=true,        
 pdfmenubar=true,        
 pdffitwindow=false,     
 pdfstartview={FitH},    
 pdftitle={My title},    
 pdfauthor={Author},     
 pdfsubject={Subject},   
 pdfcreator={Creator},   
 pdfproducer={Producer}, 
 pdfkeywords={keyword1} {key2} {key3}, 
 pdfnewwindow=true,      
 colorlinks=true,       
 linkcolor=red,          
 citecolor=cyan,        
 filecolor=magenta,      
 urlcolor=darkgreen,           
 linktocpage=true
}

\newcommand{\PF}{\mathop{\mathrm{PF}}_{B=0}}

\allowdisplaybreaks

\DeclareSymbolFontAlphabet{\mathrsfs}{rsfs}
\DeclareMathAlphabet{\mathcal}{OMS}{cmsy}{m}{n}

\newcommand{\dd}{\mathrm{d}}
 
\newcommand{\de}{\mathrm{e}}

\begin{document}
	
\title{The Quadrupole Moment of Compact Binaries \\to the Fourth post-Newtonian Order\\II. Dimensional Regularization and Renormalization}

\author{Fran\c{c}ois \textsc{Larrouturou}}\email{francois.larrouturou@iap.fr}
\affiliation{$\mathcal{G}\mathbb{R}\varepsilon{\mathbb{C}}\mathcal{O}$, Institut d'Astrophysique de Paris, UMR 7095, CNRS, Sorbonne Universit{\'e}, 98\textsuperscript{bis} boulevard Arago, 75014 Paris, France}
\affiliation{Deutsches Elektronen-Synchrotron DESY, Notkestr. 85, 22607 Hamburg, Germany}

\author{Luc \textsc{Blanchet}}\email{luc.blanchet@iap.fr}
\affiliation{$\mathcal{G}\mathbb{R}\varepsilon{\mathbb{C}}\mathcal{O}$, Institut d'Astrophysique de Paris, UMR 7095, CNRS, Sorbonne Universit{\'e}, 98\textsuperscript{bis} boulevard Arago, 75014 Paris, France}

\author{Quentin \textsc{Henry}}\email{henry@iap.fr}
\affiliation{$\mathcal{G}\mathbb{R}\varepsilon{\mathbb{C}}\mathcal{O}$, Institut d'Astrophysique de Paris, UMR 7095, CNRS, Sorbonne Universit{\'e}, 98\textsuperscript{bis} boulevard Arago, 75014 Paris, France}
\affiliation{Max-Planck-Institute for Gravitational Physics (Albert-Einstein-Institute),
Am M\"{u}hlenberg 1, 14476 Potsdam, Germany}

\author{Guillaume \textsc{Faye}}\email{faye@iap.fr}
\affiliation{$\mathcal{G}\mathbb{R}\varepsilon{\mathbb{C}}\mathcal{O}$, Institut d'Astrophysique de Paris, UMR 7095, CNRS, Sorbonne Universit{\'e}, 98\textsuperscript{bis} boulevard Arago, 75014 Paris, France}

\date{\today}

\begin{abstract} 
The regularization and renormalization of the radiative mass-type quadrupole moment of inspiralling compact binaries (without spins) is investigated at the fourth post-Newtonian (4PN) approximation of general relativity. As clear from the conservative 4PN equations of motion, a dimensional regularization has to be implemented in order to properly treat the non-linear interactions experienced by gravitational waves during their propagation toward future null infinity. By implementing such procedure, we show that the poles coming from the source moment (computed in a companion paper) are exactly cancelled in the radiative moment, as expected for a physical quantity. We thus define and obtain a ``\textit{renormalized}'' source quadrupole, three-dimensional by nature, which is an important step towards the computation of the gravitational-wave flux with 4PN accuracy. Furthermore, we explicitly prove the equivalence between the dimensional regularization and the previously used Hadamard partie finie scheme up to the 3PN order.
\end{abstract}

\pacs{04.25.Nx, 04.30.-w, 97.60.Jd, 97.60.Lf}

\maketitle

\section{Introduction}
\label{sec:introduction}

The mass quadrupole moment of compact binary systems is a crucial ingredient toward the definition of accurate gravitational wave (GW) templates in the post-Newtonian (PN) approximation. Such PN templates represent the main technique for detection and analysis of binary neutron star signals in the current network of detectors on ground, and they are at the basis of effective phenomenological methods such as EOB (effective-one-body) and IMR (inspiral-merger-ringdown), able to describe the late inspiral and merger of binary black holes (see more details in the reviews~\cite{Maggiore, BlanchetLR, BuonSathya15,Porto16}). 

Previous computations of the quadrupole moment in the case of non-spinning compact binaries achieved the 2PN order~\cite{BDI95, BDIWW95, WW96,LMRY19}, then, the 3PN order~\cite{BIJ02, BI04mult, BDEI04, BDEI05dr}, and the state-of-the-art is currently the 4PN order. A preliminary calculation was done at the 4PN order~\cite{MHLMFB20} where all the terms were computed using a series of techniques (extensively documented in~\cite{MHLMFB20}); however, as recognized in this paper, the computation was incomplete because the infra-red (IR) divergences appearing at the 4PN order were regularized thanks to the Hadamard partie finie regularization instead of the required dimensional regularization. By contrast, the ultra-violet (UV) divergences, due to the point-like nature of the source (model of compact objects by point masses without internal structure) were correctly treated with dimensional regularization~\cite{MHLMFB20}.

Recently, we analyzed in Ref.~\cite{DDR_source} the IR divergences of the (source type) mass quadrupole moment of compact binaries at 4PN order by means of dimensional regularization. This study led to the presence of poles in the dimension, \textit{i.e.} $\propto \varepsilon^{-1}\equiv (d-3)^{-1}$, arising already from the 3PN order, and, of course, also contributing at the 4PN order. The thorough computation of IR poles and the finite terms following the poles was done in~\cite{DDR_source}.

Furthermore, we also obtained~\cite{DDR_source} the contribution of propagating GW tails in the source quadrupole itself. This effect is due to retarded correlations over arbitrarily large time spans in the dynamics of the source~\cite{BD88}. It is exactly the analogue of the GW tail effect which adds a non-local piece to the conservative dynamics (equations of motion and Lagrangian/Hamiltonian) at the 4PN order~\cite{DJS14, DJS16, BBBFMa, BBBFMc, MBBF17, FStail, GLPR16, FS19, FPRS19} and beyond~\cite{Blumlein21, bini2021radiative}. As such, it induces a non-locality in time in the mass quadrupole moment at the 4PN order.

The present work is the follow-up of the companion paper~\cite{DDR_source}. We prove that the IR poles are cancelled by UV poles coming from the contributions of non-linear effects, essentially the ``tails-of-tails'' and the ``tails-of-memory'', arising in the radiative type quadrupole moment, which constitutes the physical observable at future null infinity. 

More precisely, at the 3PN order, the non-linearities are made of the tail-of-tails due to cubic interaction between two masses and the quadrupole~\cite{B98tail}, and the interactions between one mass and two mass dipoles (the latter naturally vanishes in the center-of-mass frame). Taking into account the pole term and the finite term beyond the pole, we find that the contributions from dimensional regularization in the radiative moment exactly cancel out at this order. This proves that at 3PN order the quadrupole moment can be computed using the Hadamard regularization for the IR divergences (but the dimensional regularization for the UV) and, therefore, we totally confirm the previous calculations of the 3PN quadrupole moment~\cite{BIJ02, BI04mult, BDEI04, BDEI05dr}.

Furthermore, we prove that the same cancellation holds for the 3PN current quadrupole moment, thus supporting the recent computation of this moment at 3PN order using the Hadamard regularization for the IR~\cite{HFB_courant}, and for the 3PN mass octupole moment, also confirming the previous calculation done in~\cite{FBI15}.

Finally, we demonstrate that all the poles cancel in the radiative type quadrupole moment up to 4PN order. At the 4PN order, in addition to tails-of-tails, the main non-linear effects to consider in the radiative moment are the so-called tails-of-memory, which are cubic interations between one mass monopole and two quadrupole moments. However, at the 4PN order, the net contribution of the dimensional regularization is not vanishing and, thus, we determine some finite contributions beyond the poles which are crucial to include into the final expression of the mass quadrupole moment. 

To control the occurrence of poles from the non-linear multipole interactions, we have to obtain a general expression for the retarded integral of some extended source in $d$ dimensions up to cubic order, and compute from it the difference up to order $\mathcal{O}(\varepsilon)$ between the dimensional regularization or, more precisely, the so-called $B\varepsilon$ regularization which we systematically employ~\cite{BBBFMc, MBBF17, DDR_source}, and the previously used Hadamard regularization.

The plan of this paper is as follows. We first present the derivations of the main technical ingredients used for the analysis done in this work: the generic solution of the $d$-dimensional wave equation in Sec.~\ref{sec:retint}, and the difference induced by the change of regularization scheme in Sec.~\ref{sec:diff}. Sec.~\ref{sec:MPM} then sketches the MPM (Multipolar-post-Minkowskian) method used to concretely implement the computation of this difference. The results at 3PN order are displayed in Sec.\ref{sec:reg3PN}, together with the proof that, at this order, both regularization schemes are equivalent. Finally, Sec.~\ref{sec:renorm4PN} displays the 4PN results, introduces our notion of ``renormalized mass quadrupole'' and presents its expression on quasi-circular orbits. Appendix~\ref{app:limit3d} contains the proof that the generic solution of the $d$-dimensional wave equation correctly reduces to the known results in the three-dimensional limit.

\section{The retarded integral of a multipolar extended source}
\label{sec:retint}
   
In this technical section we provide an explicit expression for the retarded solution of the wave equation in any $d$ space dimensions, in the case where the non-compact (extended) source term has a definite multipolarity. This object is the core of the iteration presented in Sec.~\ref{sec:MPM}, that we use to compute the difference between regularizations in non-linear interactions entering the gravitational wave propagation. We thus seek for the generic solution $h(\mathbf{x},t)$ solving the wave equation
\begin{equation}\label{eq:waveeq}
\Box h(\mathbf{x},t) = N(\mathbf{x},t)\,,
\end{equation}
where $\Box\equiv\Box_\eta$ is the flat d'Alembertian operator and $N(\mathbf{x},t)$ is some (non-compact support) source term. While generally the solution is provided in the Fourier domain, here we work in the physical space. The retarded solution of Eq.~\eqref{eq:waveeq} in the real domain reads (see \textit{e.g.}~\cite{BBBFMc})\footnote{We pose $c=1$ in Secs.~\ref{sec:retint}, \ref{sec:diff} and App.~\ref{app:limit3d}.}
\begin{align}\label{eq:retsol}
h(\mathbf{x},t) = - \frac{\tilde{k}}{4\pi}
\int_1^{+\infty} \!\!\dd z \,\gamma_{\frac{1-d}{2}}(z) \int
\dd^d\mathbf{x}'
\,\frac{N(\mathbf{x}',t-z\vert\mathbf{x}-\mathbf{x}'\vert)}
{\vert\mathbf{x}-\mathbf{x}'\vert^{d-2}}\,,
\end{align}
where $\tilde{k}=\Gamma(\frac{d}{2}-1)/\pi^{\frac{d}{2}-1}$ is the constant entering the Green's function of the Laplace operator, \textit{i.e.}, $\Delta\bigl(\tilde{k}\,r^{2-d}\bigr)=-4\pi\delta(\mathbf{x})$, $\Gamma$ is the Eulerian function, and we have posed
\begin{align}\label{eq:gamma}
\gamma_{\frac{1-d}{2}}(z) = \frac{2\sqrt{\pi}}{\Gamma(\frac{3-d}{2})\Gamma(\frac{d}{2}-1)}
\,\big(z^2-1\bigr)^{\frac{1-d}{2}}\,,
\end{align}
whith chosen normalization such that $\int_1^{+\infty} \dd z \,\gamma_{\frac{1-d}{2}}(z) = 1$ and $\gamma_{-1}(z)=\delta(z-1)$.

We split the spatial integration over the source point $\mathbf{x}'$ in~\eqref{eq:retsol} into an inner domain corresponding to $r'<r$, where we denote $r'\equiv\vert\mathbf{x}'\vert$ and $r\equiv\vert\mathbf{x}\vert$ with $\mathbf{x}$ being the field point, and the outer domain for which $r'>r$. Consider for instance the integration over the inner domain $r'<r$, say
\begin{align}\label{eq:retsol<}
h^{<} \equiv - \frac{\tilde{k}}{4\pi}
\int_1^{+\infty} \!\!\dd z \,\gamma_{\frac{1-d}{2}}(z) \int_{r'<r}
\dd^d\mathbf{x}'
\,\frac{N(\mathbf{x}',t-z\vert\mathbf{x}-\mathbf{x}'\vert)}
{\vert\mathbf{x}-\mathbf{x}'\vert^{d-2}}\,.
\end{align}
We obtain the multipole expansion in this domain by applying the formal Taylor expansion when $r'\to 0$. The symmetric-trace-free (STF) form of that expansion reads
\begin{align}\label{eq:hdef}
\frac{N(\mathbf{x}',t-z\vert\mathbf{x}-\mathbf{x}'\vert)}{\vert\mathbf{x}-\mathbf{x}'\vert^{d-2}} = \sum_{m=0}^{+\infty}\frac{(-)^m}{m!}\sum_{j=0}^{+\infty}\frac{\Gamma\left(\frac{d}{2}+m\right)}{\Gamma\left(\frac{d}{2}+m+j\right)}\frac{\hat{n}'_M{r'}^{2j+m}}{2^{2j}j!}\hat{\partial}_M\Delta^j\!\left[\frac{N(\mathbf{x}',t-z r)}{r^{d-2}}\right]\,,
\end{align}
where $M$ denotes a running multi-index $j_1\cdots j_m$ with $m$ indices, $\hat{n}'_M\equiv \text{STF}(n'_{j_1}\cdots n'_{j_m})$ is the STF product of $m$ unit vectors $n'_j\equiv x'_j/r'$, and we denote $\hat{\partial}_M\equiv\text{STF}(\partial_{j_1}\cdots \partial_{j_m})$ where $\partial_j\equiv\partial/\partial x^j$. Note that $\hat{\partial}_M$ and $\Delta^j$ in~\eqref{eq:hdef} act on $\mathbf{x}$, but $\mathbf{x}'$ is just spectator.

We consider the above general formula in the case where the source term $N(\mathbf{x},t)$ itself has a definite multipolarity $\ell$, \textit{i.e.} is of the type
\begin{equation}\label{eq:NL}
N(\mathbf{x},t) \equiv \hat{n}_L \,N(r,t)\,,
\end{equation}
where $L$ denotes the multi-index $i_1\cdots i_\ell$ and $\hat{n}_L\equiv \text{STF}(n_{i_1}\cdots n_{i_\ell})$ with $n_i\equiv x_i/r$. Clearly, after summing up all the multipolar pieces, while assuming the convergence of the multipolar series, there is no restriction on the generality of the solution. For a source term of the type~\eqref{eq:NL} the angular integration in the volume element $\dd^d\mathbf{x}'\equiv {r'}^{d-1}\dd{r'}\dd\Omega'_{d-1}$ can be evaluated in closed form using
\begin{equation}\label{eq:angint}
\int\dd\Omega'_{d-1} \hat{n}'_L \,\hat{n}'_M = \frac{\ell!}{2^{\ell-1}}\,\frac{\pi^{\frac{d}{2}}}{\Gamma\left(\frac{d}{2}+\ell\right)}\,\delta_{LM}\,,
\end{equation}
where $\delta_{LM}\equiv\delta_{\ell m}\delta_{(i_1\underline{j}{}_1}\cdots\delta_{i_\ell) j_\ell}$ (symmetrization over the indices $L=i_1\cdots i_\ell$); we recall the volume of the sphere $\Omega_{d-1}=2\pi^{\frac{d}{2}}/\Gamma(\frac{d}{2})$. After angular integration, the part of the integral over the domain $r'<r$, let us now call it $h_L^{<}$ since it depends on the multi-index $L$, becomes
\begin{align}\label{eq:h<}
h_L^{<} = \frac{1}{(-2)^{\ell+1}}\sum_{j=0}^{+\infty}\frac{\Gamma\left(\frac{d}{2}-1\right)}{\Gamma\left(\frac{d}{2}+\ell+j\right)}\frac{1}{2^{2j}j!}\int_1^{+\infty} \!\!\dd z \,\gamma_{\frac{1-d}{2}}(z)\int_0^r \dd r' r'^{B+2j+\ell+d-1}\hat{\partial}_L\!\left[\frac{N^{(2j)}(r',t-z r)}{r^{d-2}}\right]\,.
\end{align}
Notice that in Eq.~\eqref{eq:h<} the action of the iterated Laplacian $\Delta^j$ has reduced to $2j$ time derivatives of the source term, as indicated by the superscript in $N^{(2j)}$. Furthermore we dispose of the following useful lemma (already employed in~\cite{BBBFMc} but not given there)
\begin{align}\label{eq:lemma}
\int_1^{+\infty} \!\!\dd z \,\gamma_{\frac{1-d}{2}}(z) \,\hat{\partial}_L\!\left(\frac{F(t-z r)}{r^{d-2}}\right) = (-2)^\ell\frac{\Gamma\left(\frac{d}{2}+\ell-1\right)}{\Gamma\left(\frac{d}{2}-1\right)}\frac{\hat{n}_L}{r^{\ell+d-2}}\int_1^{+\infty} \!\!\dd z \,\gamma_{\frac{1-d}{2}-\ell}(z) F(t-z r)\,,
\end{align}
which permits to rewrite~\eqref{eq:h<} into the interesting alternative form
\begin{align}\label{eq:h<alt}
h_L^{<} = - \frac{1}{2} \sum_{j=0}^{+\infty}\frac{\Gamma\left(\frac{d}{2}+\ell-1\right)}{\Gamma\left(\frac{d}{2}+\ell+j\right)}\frac{1}{2^{2j}j!}\frac{\hat{n}_L}{r^{\ell+d-2}}
\int_1^{+\infty} \!\!\dd z \,\gamma_{\frac{1-d}{2}-\ell}(z) \int_0^r \dd {r'} {r'}^{B+\ell+2j+d-1} N^{(2j)}({r'}, t - z r)\,.
\end{align}

Very important in our approach, we use the ``$B\varepsilon$'' regularization scheme~\cite{BBBFMc, MBBF17, DDR_source}, which is a variant of the dimensional regularization (with $\varepsilon=d-3$), in which a regularization factor $(r/r_0)^B$ is inserted first in order to protect against the usual divergence of the multipole expansion when $r\to 0$, where $r_0$ is an arbitrary constant. Such factor $(r/r_0)^B$ is the same as the one used in many previous works in 3 dimensions, and here it is also used on the ``top'' of dimensional regularization. This procedure is not an \textit{ad hoc} additional regularization scheme, but comes from the matching between near zone and exterior zone and is crucial for the proper definition of the MPM algorithm in three dimensions~\cite{B98mult}. The calculations of the tail sector in the conservative equations of motion~\cite{BBBFMc, MBBF17} have indicated that the $(r/r_0)^B$ factor should be kept even in $d$ dimensions, at least in intermediate calculations. Therefore, in Eqs.~\eqref{eq:h<} and~\eqref{eq:h<alt} we have replaced the source term by $N(r,t)\longrightarrow r^B N(r,t)$. For convenience we pose $r_0=1$ but we will restore the constant $r_0$ (as well as the dimensional regularization length scale $\ell_0$) in our final results. Following the $B\varepsilon$ regularization a finite part (or Partie Finie PF~\cite{Hadamard}) at $B=0$ is always understood before considering the limit $\varepsilon\to 0$. We expect that the finite part at $B=0$ just reduces to a finite limit when $B\to 0$.

The result~\eqref{eq:h<}--\eqref{eq:h<alt} is still in the form of a formal series, and we now show how this series can be re-summed so that we obtain the corresponding ``exact'' result. To this end we use the expression of the near-zone (or PN) expansion of an antisymmetric multipolar solution of the wave equation, \textit{i.e.} of the type retarded minus advanced. In the monopolar case ($\ell=0$) this expansion is given by Eq.~(A13) of~\cite{BBBFMc}, and we see that in $d$ dimensions it involves a non-local integral. See also the discussion in Sec.~\ref{sec:MPM} and the PN expansion given by~\eqref{eq:source_dev}--\eqref{eq:source_devAB}. Let us rewrite Eq.~(A13) of~\cite{BBBFMc} in the form (with $\varepsilon=d-3$)
\begin{align}\label{eq:expmonopole}
\frac{\tilde{k}}{r^{d-2}} \int_1^{+\infty} \!\!\dd y \,\gamma_{\frac{1-d}{2}}(y) \Bigl[\widehat{F}_\varepsilon(t- y r) - \widehat{F}_\varepsilon(t+ y r)\Bigr] = -\frac{\pi^{2-\frac{d}{2}}}{2^{d-3}}\,\sum_{j=0}^{+\infty}\frac{r^{2j}}{2^{2j}j!}\,\frac{F^{(2j+1)}(t)}{\Gamma\left(\frac{d}{2}+j\right)}\,.
\end{align}
For any smooth function $F(t)$ with compact support on $\mathbb{R}$, we have introduced its ``$\varepsilon$\textit{-transform}'' $\widehat{F}_{\varepsilon}(t)$ (also smooth and with compact support) to be the function defined by analytic continuation in $\varepsilon$ as
\begin{align}\label{eq:eps-transform}
\widehat{F}_\varepsilon(t) = \frac{1}{2\cos(\frac{\pi\varepsilon}{2})\Gamma(\varepsilon)} \int_{-\infty}^{+\infty}\dd \tau \,\vert\tau\vert^{\varepsilon-1} \,F(t+\tau)\,.
\end{align}
This $\varepsilon$-transform satisfies two remarkable properties.
\begin{itemize}
\item The main one is that its inverse transform is given by the same expression but corresponding to the parameter $-\varepsilon$, thus
\begin{align}\label{eq:eps-transforminv}
F(t) = \frac{1}{2\cos(\frac{\pi\varepsilon}{2})\Gamma(-\varepsilon)} \int_{-\infty}^{+\infty}\dd \tau \,\vert\tau\vert^{-\varepsilon-1} \,\widehat{F}_\varepsilon(t+\tau)\,.
\end{align}
This nice property can easily be proved by going to the Fourier domain, using \textit{e.g.}~(3.17)--(3.18) in~\cite{BBBFMc}.\footnote{Hence we can check that Eq.~\eqref{eq:expmonopole} is indeed equivalent to Eq.~(A13) of~\cite{BBBFMc}.} 
\item The second important property of the $\varepsilon$-transform is that it reduces to the identity in the limit when $\varepsilon\to 0$:
\begin{align}\label{eq:limiteps0}
\lim_{\varepsilon\to 0}\widehat{F}_\varepsilon(t) = F(t)\,.
\end{align}
However, the transform~\eqref{eq:eps-transform}--\eqref{eq:eps-transforminv} is valid by analytic continuation for any $\varepsilon\in\mathbb{C}$.
\end{itemize}

Next we obtain the expansion for any multipolar antisymmetric wave by applying on Eq.~\eqref{eq:expmonopole} the STF multi-spatial derivative operator $\hat{\partial}_L$, and we obtain\footnote{Using the fact that $\hat{\partial}_L r^{2j} = \frac{2^\ell j!}{(j-\ell)!}\,\hat{n}_L \,r^{2j-\ell}$ when $j\geqslant\ell$ and is zero when $0\leqslant j\leqslant\ell-1$.} 
\begin{align}\label{eq:}
\tilde{k} \int_1^{+\infty} \!\!\dd y \,\gamma_{\frac{1-d}{2}}(y) \,\hat{\partial}_L\Biggl[\frac{\widehat{F}_\varepsilon(t- y r) - \widehat{F}_\varepsilon(t+ y r)}{r^{d-2}}\Biggr] = -\frac{\pi^{2-\frac{d}{2}}}{2^{\ell+d-3}}\,\sum_{j=0}^{+\infty}\frac{\hat{x}_L r^{2j}}{2^{2j}j!}\,\frac{F^{(2j+2\ell+1)}(t)}{\Gamma\left(\frac{d}{2}+\ell+j\right)}\,.
\end{align}
Thanks to the relation~\eqref{eq:expmonopole} with the substitution $d \rightarrow d+2\ell$, one can finally re-express the result for $h_L^<$ as given by~\eqref{eq:h<} in the following closed-analytic ``exact'' form
\begin{align}\label{eq:h<exact}
h_L^{<} &= \tilde{k} \,\Gamma_\ell \int_1^{+\infty} \dd y \,\gamma_{\frac{1-d}{2}}(y)\int_1^{+\infty} \dd z \,\gamma_{\frac{1-d}{2}-\ell}(z)\\&\qquad\qquad\times\int_0^r\!\dd {r'} \,{r'}^{B-\ell+1}\hat{\partial}_L\!\left[\frac{\widehat{N}_\varepsilon^{(-2\ell-1)}\bigl({r'}, t - y r - z {r'}\bigr) - \widehat{N}_\varepsilon^{(-2\ell-1)}\bigl({r'}, t - y r + z {r'}\bigr)}{r^{d-2}}\right]\,,\nonumber
\end{align}
where the superscript $(-2\ell-1)$ indicates $2\ell+1$ time anti-derivatives\footnote{As usual in our formalism, we assume stationarity in the remote past, and anti-derivatives are defined to be those which vanish when $t\to-\infty$.} and we have posed $\Gamma_\ell \equiv 2^{\ell+d-4} \pi^{\frac{d}{2}-2}\,\Gamma(\frac{d}{2}+\ell-1)$, such that $\lim_{\varepsilon\to 0} \Gamma_\ell=\frac{(2\ell-1)!!}{2^{\ell-1}}$. We can essentially repeat the same steps of the previous calculation with only some simple adaptations, for the outer part of the solution $h_L^{>}$ corresponding to $r'>r$, and we obtain
\begin{align}\label{eq:h>exact}
h_L^{>} &= \tilde{k} \,\Gamma_\ell \int_1^{+\infty} \dd y \,\gamma_{\frac{1-d}{2}}(y)\int_1^{+\infty} \dd z \,\gamma_{\frac{1-d}{2}-\ell}(z)\\&\qquad\qquad\times\int_r^{+\infty}\!\dd {r'} \,{r'}^{B-\ell+1}\hat{\partial}_L\!\left[\frac{\widehat{N}_\varepsilon^{(-2\ell-1)}\bigl({r'}, t - y r - z {r'}\bigr) - \widehat{N}_\varepsilon^{(-2\ell-1)}\bigl({r'}, t + y r - z {r'}\bigr)}{r^{d-2}}\right]\,.\nonumber
\end{align}
A finite part PF when $B\to 0$ is to be understood in front of both Eqs.~\eqref{eq:h<exact}--\eqref{eq:h>exact}. Recalling the definition~\eqref{eq:eps-transform} of the $\varepsilon$-transform we see that~\eqref{eq:h<exact}--\eqref{eq:h>exact} involve three non-local integrals over $y$, $z$ and $\tau$, that are specifically due to $d$ dimensions and reduce to trivial local results in 3 dimensions. Notice that in the complete solution the first terms in~\eqref{eq:h<exact} and~\eqref{eq:h>exact} merge to form a single integral from 0 to $+\infty$. Thus we may write the complete solution $h_L = h_L^{<} + h_L^{>}$ as
\begin{align}\label{eq:hcomplet}
&h_L = \tilde{k}\,\Gamma'_\ell \,\frac{\hat{n}_L}{r^{d+\ell-2}} \int_1^{+\infty} \dd y \,\gamma_{\frac{1-d}{2}-\ell}(y)\int_1^{+\infty} \dd z \,\gamma_{\frac{1-d}{2}-\ell}(z)\\&\qquad\times\Biggl\{\int_0^{+\infty}\!\dd {r'} \,{r'}^{B-\ell+1}\widehat{N}_\varepsilon^{(-2\ell-1)}\bigl({r'}, t - y r - z {r'}\bigr)\nonumber\\&\qquad - \int_0^r\!\dd {r'} \,{r'}^{B-\ell+1}\widehat{N}_\varepsilon^{(-2\ell-1)}\bigl({r'}, t - y r + z {r'}\bigr) - \int_r^{+\infty}\!\dd {r'} \,{r'}^{B-\ell+1}\widehat{N}_\varepsilon^{(-2\ell-1)}\bigl({r'}, t + y r - z {r'}\bigr)\Biggr\}\,,\nonumber
\end{align}
where we pose $\Gamma'_\ell \equiv (-)^\ell 2^{2\ell+d-4} \pi^{\frac{d}{2}-2}\,[\Gamma(\frac{d}{2}+\ell-1)]^2/\Gamma(\frac{d}{2}-1)$, and we have chosen to apply the lemma~\eqref{eq:lemma} on the $y$ integration in order to express the result. Note that the first term in~\eqref{eq:hcomplet} is an homogeneous retarded solution of the wave equation in $d$ dimensions. In the Appendix~\ref{app:limit3d}, we explicitly check the limit of our result~\eqref{eq:hcomplet} in three dimensions, and we gladly recover a formula derived previously in~\cite{BD86}.

We end up this section by two remarks. For the computation of the conservative Lagrangian and equations of motion at 4PN order, and handling the conservative part of the tail effect at that order, it was shown in Ref.~\cite{BBBFMc} that the relevant quantity to be considered is $h_L^0$, defined by $h_L^{>}$ as given by~\eqref{eq:h>exact} but in which the radial integral from $r$ to $+\infty$ is replaced by the integral from 0 to $+\infty$:\footnote{This solution is given in expanded form by Eq.~(3.20) in~\cite{BBBFMc}.}
\begin{align}\label{eq:h0}
h_L^0 &= \tilde{k} \,\Gamma_\ell \int_1^{+\infty} \dd y \,\gamma_{\frac{1-d}{2}}(y)\int_1^{+\infty} \dd z \,\gamma_{\frac{1-d}{2}-\ell}(z)\\&\qquad\qquad\times\int_0^{+\infty}\!\dd {r'} \,{r'}^{B-\ell+1}\hat{\partial}_L\!\left[\frac{\widehat{N}_\varepsilon^{(-2\ell-1)}\bigl({r'}, t - y r - z {r'}\bigr) - \widehat{N}_\varepsilon^{(-2\ell-1)}\bigl({r'}, t + y r - z {r'}\bigr)}{r^{d-2}}\right]\,.\nonumber
\end{align}
Indeed this $h_L^0$ is an homogeneous solution of the wave equation, $\Box h_L^0 = 0$, and furthermore is regular when $r\to 0$, \textit{i.e.} is of the type retarded minus advanced. Therefore it was proved that this solution can be extended inside the source by matching, and it permitted the computation of the 4PN tail effect in the conservative equations of motion (with result in perfect agreement with the one derived by the effective-field-theory approach~\cite{GLPR16}). 

Similarly we will find in the next section that in order to compute the contribution from tails in the far zone, it is sufficient to define $h_L^\infty$ by $h_L^{<}$ as given by~\eqref{eq:h<exact}, but in which the radial integral from 0 to $r$ is replaced by the integral from 0 to $+\infty$:
\begin{align}\label{eq:hinf}
h_L^\infty &= \tilde{k} \,\Gamma_\ell \int_1^{+\infty} \dd y \,\gamma_{\frac{1-d}{2}}(y)\int_1^{+\infty} \dd z \,\gamma_{\frac{1-d}{2}-\ell}(z)\\&\qquad\qquad\times\int_0^{+\infty}\!\dd {r'} \,{r'}^{B-\ell+1}\hat{\partial}_L\!\left[\frac{\widehat{N}_\varepsilon^{(-2\ell-1)}\bigl({r'}, t - y r - z {r'}\bigr) - \widehat{N}_\varepsilon^{(-2\ell-1)}\bigl({r'}, t - y r + z {r'}\bigr)}{r^{d-2}}\right]\,.\nonumber
\end{align}
Again this is an homogeneous solution of the wave equation, $\Box h_L^\infty = 0$, but this time a retarded solution, therefore regular at infinity, when $r\to +\infty$. This is that solution $h_L^\infty$ which will be responsible for the contribution of far zone tails in the difference between the dimensional and Hadamard regularizations, as we will show in the next section.

\section{The difference between two regularization schemes}
\label{sec:diff}

Up to vanishingly small terms when $\varepsilon\to 0$, the difference between the dimensional regularization and the Hadamard one comes only from the bound ${r'}=0$ in the general formula for the retarded integral, \textit{i.e.} it comes only from the part of the integral $h_L^{<}$ given by~\eqref{eq:h<exact}. This is clear because only the bound ${r'}=0$ can give a pole in dimensional regularization, and any integration over an interval excluding ${r'}=0$ will have a finite limit when $\varepsilon\to 0$, and so cancels out between the two regularization in the limit $\varepsilon\to 0$. This further means that for the computation of the difference we can replace the integral from 0 to $r$ in~\eqref{eq:h<exact} by the integral from 0 up to $+\infty$ (or any finite constant), \textit{i.e.} we can just consider the homogeneous retarded solution $h_L^\infty$ given by~\eqref{eq:hinf}.

Thus our statement is that modulo terms of order $\varepsilon$ the difference is
\begin{align}\label{eq:DhL}
\mathcal{D} h_L = \mathcal{D} h^\infty_L + \mathcal{O}(\varepsilon)\,,
\end{align}
where the difference operator $\mathcal{D}$ is formally defined as the commutator between the three-dimensional limit (\textit{viz.} $\lim_{3d}$) and the PF operator:
\begin{equation}\label{eq:Ddef}
\mathcal{D} A \equiv
\left[\lim_{3d}\,,\PF\right]\,A
= 
\lim_{3d}\left\lbrace\PF\,A\right\rbrace - \PF \left\lbrace\lim_{3d}\,A\right\rbrace 
=
 A^{B\varepsilon} - A^\text{Had} \,.
\end{equation}
Here, by ``3d'' limit, we really mean the expansion $\varepsilon\to 0$, while crucially keeping all the divergent poles $\varepsilon^{-1}$, and neglecting terms $\mathcal{O}(\varepsilon)$. By contrast, $\PF$ discards the poles $B^{-1}$, so the difference $\mathcal{D}A$ is actually a function of the poles $\varepsilon^{-1}$ but does not depend on $B$. 

Since $h^\infty_L$ is a retarded homogeneous solution of the wave equation in $d$ dimensions, it can be written in multipolar form as
\begin{align}\label{eq:hLinfdef}
h_L^{\infty} = \hat{\partial}_L\,\biggl[\frac{\tilde{k}}{r^{d-2}} \,\int_1^{+\infty} \dd y \,\gamma_{\frac{1-d}{2}}(y) \,H(t-y r)\biggr]\,,
\end{align}
where we pose
\begin{align}\label{eq:Hexpr}
H(t) &= \Gamma_\ell \int_1^{+\infty} \dd z \,\gamma_{\frac{1-d}{2}-\ell}(z)\int_0^{+\infty}\!\dd {r'} \,{r'}^{B-\ell+1}\,\left[\widehat{N}_\varepsilon^{(-2\ell-1)}\bigl({r'}, t - z {r'}\bigr) - \widehat{N}_\varepsilon^{(-2\ell-1)}\bigl({r'}, t + z {r'}\bigr)\right]\,.
\end{align}
We recall that the ``$\varepsilon$-transform'' operation is defined by~\eqref{eq:eps-transform}.

As we consider only the limit ${r'}\to 0$ we insert into~\eqref{eq:Hexpr} the near-zone expansion of the source term $N(r,t)$ in $d$ dimensions, which will turn out to be always of the generic type (as shown in Sec.~\ref{sec:MPM})
\begin{align}\label{eq:Ndexp}
N(r,t) = \sum_{p,q} r^{-p-q\varepsilon} f_{p,q}(t)\,,
\end{align}
where the coefficients $f_{p,q}(t)$ may be complicated non-local integrals involving poles, and the singularity is not essential, \textit{i.e.} the maximal divergence when $r\to 0$ is finite, say $p\leqslant p_0$. Furthermore, we have explicitly checked that $q$ takes values in a finite range, say $q_0\leqslant q\leqslant q_1$. The $d$-dimensional coefficients $f_{p,q}(t)$ match the corresponding three-dimensional coefficients when $\varepsilon\to 0$, in the sense that
\begin{align}\label{eq:fpmatch}
\lim_{\varepsilon\to 0} \,\sum_q r^{-q\varepsilon}\, f_{p,q}(t) =
f^{(3d)}_{p}(t) + f^{(3d,\text{ln})}_{p}(t)\ln \left(\frac{r}{\ell_0}\right) \,.
\end{align}
The three-dimensional logarithm comes from possible poles in $f_{p,q}$, see also Eq.~(3.9) of~\cite{DDR_source}, but in our practical calculations the limit~\eqref{eq:fpmatch} is finite. Furthermore, in our case, the source does not bear double poles so that only simple logarithms show up in the three-dimensional limit. By inserting~\eqref{eq:Ndexp} into~\eqref{eq:Hexpr} we obtain
\begin{align}\label{eq:H2}
H(t) \!&=\! \sum_{p,q} \Gamma_\ell \!\int_1^{+\infty} \!\!\dd z \,\gamma_{-1-\ell-\frac{\varepsilon}{2}}(z)\!\int_0^{+\infty}\!\!\dd {r'} {r'}^{B-\ell+1-p-q\varepsilon}\,\biggl[{}_\varepsilon\widehat{f}_{p,q}^{(-2\ell-1)}\bigl(t - z {r'}\bigr) \!-\! {}_\varepsilon\widehat{f}_{p,q}^{(-2\ell-1)}\bigl(t + z {r'}\bigr)\biggr]\,,
\end{align}
and by posing $\rho = z r'$, we find that the integral over $z$ factorizes out and can nicely be computed in closed analytic form.\footnote{For that we use $$\int_1^{+\infty} \dd z \,z^{\alpha} \,\gamma_{s}(z) = \sqrt{\pi}\frac{\Gamma(-s-\frac{\alpha+1}{2})}{\Gamma(-s-\frac{1}{2})\Gamma(\frac{1-\alpha}{2})}.$$} Finally, we are able to perform a series of integrations by parts (with all integrated terms at infinity vanishing) to arrive at
\begin{align}\label{eq:H}
H(t) \!&=\! - 2^{\ell-1+\varepsilon} \pi^{\frac{\varepsilon}{2}} \sum_{p,q} \int_0^{+\infty}\!\!\dd \rho \,Q_\ell^{p,q}\bigl(B,\varepsilon,\rho\bigr)\,\biggl[(-)^{\ell+p}{}_\varepsilon\widehat{f}_{p,q}^{(-\ell-2+p)}\bigl(t - \rho\bigr) + {}_\varepsilon\widehat{f}_{p,q}^{(-\ell-2+p)}\bigl(t + \rho\bigr)\biggr]\,,
\end{align}
in which we have inserted $\Gamma_\ell \equiv 2^{\ell+d-4} \pi^{\frac{d}{2}-2}\Gamma(\frac{d}{2}+\ell-1)$, and posed
\begin{align}\label{eq:Q}
Q_\ell^{p,q}\bigl(B,\varepsilon,\rho\bigr) \equiv \frac{\Gamma\left(\frac{B+\ell+3-p-(q-1)\varepsilon}{2}\right)\Gamma\left(\text{$-B+q\varepsilon$}\right)}{\Gamma\left(\frac{B-\ell+3-p-q\varepsilon}{2}\right)\Gamma\left(-B+\ell-1+p+q\varepsilon\right)}\,\rho^{B-q\varepsilon}\,.
\end{align}

With this result we straightforwardly obtain the difference $\mathcal{D}H$ corresponding to the calculation with the $B\varepsilon$ method minus the Hadamard regularization which is just the finite part at $B=0$ in three dimensions. Of course we use the link between $d$ dimensions and 3 dimensions saying that the functions $f_{p,q}(t)$ match their counterparts in 3 dimensions, see~\eqref{eq:fpmatch}. From~\eqref{eq:H}, it is clear that the difference reads as
\begin{align}\label{eq:DH_gen}
\mathcal{D}H(t) &= - 2^{\ell-1+\varepsilon}\pi^{\frac{\varepsilon}{2}} \sum_{p,q} \int_0^{+\infty}\!\dd \rho \,\mathcal{D}Q_\ell^{p,q}\bigl(\varepsilon,\rho\bigr)\,\biggl[(-)^{\ell+p}{}_\varepsilon\widehat{f}_{p,q}^{(-\ell-2+p)}\bigl(t - \rho\bigr) + {}_\varepsilon\widehat{f}_{p,q}^{(-\ell-2+p)}\bigl(t + \rho\bigr)\biggr]\,,
\end{align}
where we recall that $\mathcal{D}Q_\ell^{p,q}\bigl(\varepsilon,\rho\bigr)$ denotes the difference of the quantities~\eqref{eq:Q} evaluated with the two regularizations, and given explicitly by~\eqref{eq:Ddef} as
\begin{align}\label{eq:DQ}
\mathcal{D}Q_\ell^{p,q}\bigl(\varepsilon,\rho\bigr)\equiv \lim_{3d}\left\{\PF \bigl[Q_\ell^{p,q}\bigl(B,\varepsilon,\rho\bigr)\bigr]\right\} - \PF \bigl[Q_\ell^{p,q}\bigl(B,0,\rho\bigr)\bigr]\,.
\end{align}
Finally it is a straightforward (although fairly tedious) matter to evaluate explicitly the difference~\eqref{eq:DQ} neglecting vanishing terms in the limit $\varepsilon\to 0$. We find that there are two cases where the difference is not zero:
\begin{enumerate}
\item When $p=\ell+2i+3$ where $i\in\mathbb{N}$ and $q\neq 1$, in which case we have
\begin{align}\label{eq:DQ_pl3}
\hspace{-0.3cm}\mathcal{D}Q_\ell^{p,q} = \frac{(-)^\ell(\ell+i)!}{i!(2\ell+2i+1)!}\frac{1}{q-1}\biggl[\frac{1}{\varepsilon} - \sum_{k=0}^{\ell+i}\frac{1}{2k+1} - \frac{1}{2}\gamma_\text{E} - \ln\left(\frac{\rho \,r_0^{q-1}}{\ell_0^q}\right)\biggr] + \mathcal{O}\left(\varepsilon\right)\,.
\end{align}
Note the important point that the exclusion of the $q=1$ is not an artificial requirement to ensure that the formula is well-defined. Indeed, when $q=1$ the difference $\mathcal{D}Q_\ell^{p,1}$ is exactly vanishing; this point will be of uttermost importance in Sec.~\ref{sec:MPM}.
\item When $p=-\ell+2i+2$ where $i\in\mathbb{N}$ and $q\neq 0$, where we have
\begin{align}\label{eq:DQ_autre}
\hspace{-0.3cm}\mathcal{D}Q_\ell^{p,q} = (-)^i\frac{(2\ell-2i-1)!!}{2^{\ell+i} i!}\frac{1}{q}\biggl[\frac{1}{\varepsilon} + \sum_{k=0}^{|\ell-i|-1}\frac{1}{2k+1} - \frac{1}{2}\gamma_\text{E} +  \ln\left(\frac{\ell_0^q}{2r_0^q}\right)\biggr] + \mathcal{O}\left(\varepsilon\right)\,.
\end{align}
with the convention that $(-2n-1)!!(2n+1)!! \equiv (-)^n (2n+1)$ for any integer $n \in \mathbb{N}$. 
\end{enumerate}
All other cases give zero (no difference). In case 2, since~\eqref{eq:DQ_autre} does not depend on $\rho$ the difference is local, and turns out to be vanishing. Indeed, due to the even nature of $p+\ell$, the two integrals corresponding to the two terms in the square brackets of~\eqref{eq:DH_gen} exactly cancel out. On the other hand, the case 1 seems to be non-local as there is a $\ln\rho$ remaining and therefore a non-local integral associated. However we find that this non-local integral exactly cancels out the one present in the coefficients ${}_\varepsilon\widehat{f}_{p,q}$ up to order $\mathcal{O}(\varepsilon)$, \textit{cf.}~Eq.~\eqref{eq:eps-transform} of the $\varepsilon$-transform, so that the final result does not contain a remaining integral over $\rho$. Note that we have not considered the $\mathcal{O}(\varepsilon)$ term, as we have explicitly verified that none of the contributing $f_{p,q}$ (\textit{i.e.} such that $p-\ell-3 \in 2\mathbb{N}$) bears poles. If it were the case, the difference would turn out to be non-local. Fortunately this is not our case and thus we are able to write down our final result as (with $\bar{q}=4\pi \de^{\gamma_\text{E}}$)
\begin{align}\label{eq:DH}
\mathcal{D}H(t) &=(-2)^\ell \sum_{i=0}^{+\infty} \frac{(i+\ell)!}{i!(2\ell+2i+1)!} \sum_{p,q \not= 1} \frac{1}{q-1}\biggl[\frac{1}{\varepsilon} - \sum_{k=0}^{\ell+i}\frac{1}{2k+1} + \ln\left(\sqrt{\bar{q}}\,\frac{\ell_0^{q-1}}{r_0^{q-1}}\right)\biggr]\frac{f_{p,q}^{(2i)}(t)}{c^{2i}}\,,
\end{align}
plus the neglected remainder $\mathcal{O}(\varepsilon)$. It is understood that the sum over $i$ is actually finite because the singularity in the source term~\eqref{eq:Ndexp} is finite (\textit{i.e.}~$p\leqslant p_0$ for some $p_0\in\mathbb{N}$). In~\eqref{eq:DH} we have restored the relevant Hadamard regularization scale $r_0$ and dimensional regularization scale $\ell_0$. Again we emphasize that this result is local, as long as the $f_{p,q}(t)$ are themselves local. As presented in the Sec.~\ref{sec:MPM}, this is always the case, and thus the complicated difference between the two regularization schemes of non-linear interactions is local. This appears to be a non-trivial result, considering that the tails and their iterations are by essence non-local.

To summarize, we have computed the difference between regularization schemes in the metric in harmonic coordinates. From~\eqref{eq:DhL} and~\eqref{eq:hLinfdef} we have expressed this difference, for each component of the metric and each multipolarity $\ell$, see~\eqref{eq:NL}, in the form of a retarded solution of the d'Alembertian equation in $d$ dimensions,
\begin{align}\label{eq:DhL_DH}
\mathcal{D}h_L = \hat{\partial}_L\,\biggl[\frac{\tilde{k}}{r^{d-2}} \,\int_1^{+\infty} \dd y \,\gamma_{\frac{1-d}{2}}(y) \,\mathcal{D}H(t-y r)\biggr]\,,
\end{align}
where the function $\mathcal{D}H$ is given by~\eqref{eq:DH} for any source term of the form~\eqref{eq:Ndexp} 

\section{The MPM algorithm for non-linear interactions}
\label{sec:MPM}

The goal of this work is to apply the formulas derived in the previous section to the case of the non-linear interactions entering the mass quadrupole moment at 4PN. A quick dimensional analysis shows that such interactions are at most cubic in $G$. Thus, we need to compute the cubic source term entering the right-hand side of the wave equation~\eqref{eq:waveeq}. To this end, we rely on the so-called Multipolar-Post-Minkowskian (MPM) algorithm~\cite{BD86, B98mult}.

The MPM algorithm builds the sources (and solutions) of the wave equation~\eqref{eq:waveeq} in an iterative fashion, orders by orders in $G$, hence the name. 
Its starting point is the so-called canonical metric, which is a generalization in $d$ dimensions of Thorne's linearized metric~\cite{Th80}. The recent calculation in $d$ dimensions~\cite{HFB_courant} writes it as
\begin{subequations}\label{eq:hlin_MSK}
	\begin{align}
& h^{00}_\text{can\,1}=
	- \frac{4}{c^2}\sum_{\ell \geqslant 0}\frac{(-)^\ell}{\ell !} \,\hat{\partial}_L\,\widetilde{\mathcal{M}}_L\,,\\
&
h^{0i}_\text{can\,1} =
\frac{4}{c^3}\sum_{\ell \geqslant 1}\frac{(-)^\ell}{\ell !}\,\left[\hat{\partial}_{L-1}\,\widetilde{\mathcal{M}}_{iL-1}^{(1)}+\frac{\ell}{\ell+1}\hat{\partial}_L\widetilde{\mathcal{S}}_{i\vert L}\right]\,,\\
&
h^{ij}_\text{can\,1} =
- \frac{4}{c^4}\sum_{\ell \geqslant 2}\frac{(-)^\ell}{\ell !}\,\left[
\hat{\partial}_{L-2}\,\widetilde{\mathcal{M}}_{ijL-2}^{(2)}+\frac{2\ell}{\ell+1}\hat{\partial}_{L-1}\widetilde{\mathcal{S}}^{(1)}_{(i\vert j)L-1}+\frac{\ell-1}{\ell+1}\hat{\partial}_L\widetilde{\mathcal{K}}_{ij\vert L}\right]\,,
\end{align}
\end{subequations}
where we have defined the ``tilded'' versions of the moments as, \textit{e.g.},
\begin{equation}\label{eq:tildedML}
\widetilde{\mathcal{M}}_L(r,t) \equiv \frac{\tilde{k}}{r^{d-2}}\int_1^{+\infty}\!\!\dd y\,\gamma_\frac{1-d}{2}(y)\,M_L\left(t- \frac{yr}{c}\right)\,,
\end{equation}
and where $M_L$ and $S_{i\vert L}$ denote the mass and current canonical moments, together with the additional moment $K_{ij\vert L}$, which does not exist in 3 dimensions. 

We let the reader refer to~\cite{BlanchetLR} for a comprehensive review of the MPM formalism.\footnote{Note that the procedure is formally similar in three-dimensions (as presented in~\cite{BlanchetLR}) and in $d$-dimensions (as should be implemented here); however the three-dimensional moments have to be replaced by their $d$-dimensional tilded counterparts, \textit{e.g.} $M_L(t-r/c)/r \longrightarrow \widetilde{\mathcal{M}}_L(r,t)$ as defined in~\eqref{eq:tildedML}.} We will only note that, due to its iterative nature, we need the knowledge of the quadratic solution to the wave equation in order to build up the cubic source. However, let us recall that the difference between regularization schemes is a near zone effect [as only the  ${r'}\to 0$ limit of the integral~\eqref{eq:hinf} plays a role], and so we can focus on the near zone behaviour of the source only. This notably implies that we do not require the knowledge of the whole quadratic interactions to build up the cubic source, but only their near zone behaviour, which we can iterate following the MPM procedure.

We discuss the near zone expansion of a quadratic metric, taking first the example of the quadratic interaction $M\times M_{ij}$. Later we will say a word on the more difficult interaction $M_{ij} \times M_{ij}$, but the generalization to other interactions is almost straightforward. The source of a $d$-dimensional quadratic interaction like $M\times M_{ij}$ (for which the monopole $\widetilde{\mathcal{M}} = \tilde{k} M r^{2-d}$ is static) takes the generic form
\begin{equation}\label{eq:sourceterm}
N(\mathbf{x},t) = \hat{n}_L\,\frac{\ell_0^{q\varepsilon}}{r^{p+q\varepsilon}}\int_1^{+\infty}\!\!\dd z\,\gamma_\frac{1-d}{2}(z)\,z^k\,F\left(t-\frac{zr}{c}\right)\,,
\end{equation}
and $F$ involves some time derivative of $M_{ij}$. The exact prescription to be applied would naturally be the propagator~\eqref{eq:retsol}, as we discussed in details in Sec.~\ref{sec:retint}. Nevertheless, we do not need to use it for the quadratic interactions, as we only seek for the near zone behaviour. Following~\cite{BBBFMc}, the formal PN expansion of the quadratic solution is made of two terms,
\begin{align}\label{eq:hLbar}
\overline{h}_L = \text{PF}\, \Box_\text{R}^{-1}\overline{N} - \PF\sum_{m\in\mathbb{N}}\frac{(-)^m}{m!}\sum_{j\in\mathbb{N}}\Delta^{-j}\hat{x}_M\int\!\dd^d\mathbf{x}'\left(\frac{r'}{r_0}\right)^B
\hat{\partial}_M'\left[\frac{\widetilde{\mathcal{N}}^{(2j)}(\mathbf{y},t-zr'/c)}{c^{2j}\,r'{}^{d-2}}\right]_{\mathbf{y}=\mathbf{x}'}\,,
\end{align}
where the ``tilde'' operation (with respect to variables $r',t$) is defined by~\eqref{eq:tildedML} and we denoted
\begin{equation}\label{eq:DeltajxL}
\Delta^{-j}\hat{x}_L \equiv \frac{\Gamma\left(\frac{d}{2}+\ell\right)}{\Gamma\left(\frac{d}{2}+\ell+j\right)}\frac{r^{2j}\hat{x}_L}{2^{2j}j!}\,.
\end{equation}
The first term is a particular solution built up (in the sense of the PF) from the formal PN expansion $\overline{N}$ of the source term~\eqref{eq:sourceterm}. Introducing $\tau = zr/c$ and using the asymptotic expansion when $z\to+\infty$ of the function $\gamma_\frac{1-d}{2}$ given by~\eqref{eq:gamma}, \textit{i.e.}
\begin{equation}
\gamma_\frac{1-d}{2}(z) \underset{z\to+\infty}{\sim} \ \sum_{j=0}^{+\infty}\frac{(-)^j}{j!}\frac{2\sqrt\pi\,z^{-2-2j-\varepsilon}}{\Gamma\left(-j-\frac{\varepsilon}{2}\right)\Gamma\left(\frac{1+\varepsilon}{2}\right)}\,,
\end{equation}
we can decompose $\overline{N}$ into ``even'' and ``odd'' parts given by
\begin{subequations}\label{eq:source_dev}
\begin{align}
& \overline{N}^\text{even} = \sum_{j\in\mathbb{N}}A_j^k\,\frac{r^{2j+\kappa-p-q\varepsilon}}{c^{2j+\kappa}}\,F^{(2j+\kappa)}(t)\,\hat{n}_L\,,\label{eq:source_dev_even}\\
& \overline{N}^\text{odd} =\sum_{j\in\mathbb{N}}B_j^k\,\frac{r^{1+2j-k-p+(1-q)\varepsilon}}{c^{1+2j-k+\varepsilon}}\int_0^\infty\!\!\dd\tau\, \tau^{-\varepsilon}\,F^{(2j+2-k)}(t-\tau)\,\hat{n}_L\,,\label{eq:source_dev_odd}
\end{align}
\end{subequations}
where $\kappa\equiv k-2[\frac{k}{2}]$, \textit{i.e.} $\kappa = 0$ if $k$ is even and $1$ if $k$ is odd, and the coefficients that enter those expressions are
\begin{subequations}\label{eq:source_devAB}
	\begin{align}
	& A_j^k \equiv 
	\frac{(-)^j}{2^{2j+\kappa}j!}\frac{\Gamma\left(\frac{1}{2}-j-\kappa\right)}{\Gamma\left(\frac{1+\varepsilon}{2}\right)}\frac{\Gamma\left(\frac{1-\kappa-k+\varepsilon}{2}-j\right)}{\Gamma\left(\frac{1-\kappa-k}{2}-j\right)}\,,\\
	& B_j^k \equiv
	\frac{(-)^k}{2^{2j}j!}
	\frac{2\sqrt\pi\,\Gamma\left(2-k+\varepsilon\right)}{\Gamma\left(\frac{1+\varepsilon}{2}\right)\Gamma\left(1-\varepsilon\right)\Gamma\left(\frac{-\varepsilon}{2}\right)}
	\frac{\Gamma\left(1+j+\frac{\varepsilon}{2}\right)}{\Gamma\left(1+\frac{\varepsilon}{2}\right)}
	\frac{\Gamma\left(1-\frac{k}{2}+\frac{\varepsilon}{2}\right)}{\Gamma\left(1+j-\frac{k}{2}+\frac{\varepsilon}{2}\right)}
	\frac{\Gamma\left(\frac{3-k+\varepsilon}{2}\right)}{\Gamma\left(j+\frac{3-k+\varepsilon}{2}\right)}
	\,.
	\end{align}
\end{subequations}
The expressions~\eqref{eq:source_dev}--\eqref{eq:source_devAB} generalize those of~\cite{BBBFMc} in the case of a non-vanishing $k$. 

Once integrated with the PN-expanded propagator, using the ``Matthieu'' formula, \textit{i.e.} Eq.~(3.16) in~\cite{DDR_source}, we find that Eqs.~\eqref{eq:source_dev} yield poles. Fortunately, the three-dimensional limit is non-pathological, since those poles then disappear by mutual cancellation. They simply generate logarithmic terms as emphasized by~\eqref{eq:fpmatch}. Moreover, such poles do not contribute to the difference between regularization schemes in the iterated cubic solution, as already advertized (they do not enter the terms bearing $p -\ell - 3 \in 2\mathbb{N}$).

As for the homogeneous solution, which is the second term of~\eqref{eq:hLbar}, after a series of transformations following~\cite{BBBFMc}, it becomes
\begin{align}\label{eq:hhom_dev}
\overline{h}_L^\text{hom} = \frac{(-)^{p+\ell}}{d+2\ell-2} \PF \,\frac{\Gamma\left(q\varepsilon-B\right)}{\Gamma\left(p+\ell-1+q\varepsilon-B\right)}\,C^{k,p,q}_\ell\,\sum_{j\in\mathbb{N}}\Delta^{-j}\hat{x}_L\int_0^\infty\!\!\dd\tau\frac{\tau^{B-q\varepsilon}}{r_0^B}\frac{F^{(2j+\ell+p-1)}(t-\tau)}{c^{2j+\ell+p+q\varepsilon-B}}\,,
\end{align}
where we have denoted
\begin{equation}\label{eq:Clkpq}
C^{k,p,q}_\ell \equiv
\int_1^{+\infty}\!\!\dd y\,\gamma_{\frac{1-d}{2}-\ell}(y)\int_1^{+\infty}\!\!\dd z\,\gamma_{\frac{1-d}{2}}(z)\,z^k(y+z)^{\ell-2+p+q\varepsilon-B}\,.
\end{equation}
The coefficients $C^{k,p,q}_\ell$ are generalizations for $q \in \mathbb{Z}$ of the ones introduced in~\cite{BBBFMc}. They can be computed by methods similar to those described in the App. D of~\cite{BBBFMc}.

The three types of terms we have met: integrated ``even'' and ``odd'' PN expansions, as well as the homogeneous solutions, are to be inserted into the cubic source which, again, can be PN-expanded and will take the same general form as in~\eqref{eq:source_dev}, which confirms our claim on the structure of the cubic source~\eqref{eq:Ndexp}.

An important point to note is that only the ``even'' contributions~\eqref{eq:source_dev_even} will contribute to the final result~\eqref{eq:DH}. Indeed we have explicitly checked that, due to the particular powers of $r$, the ``odd'' terms given by~\eqref{eq:source_dev_odd} will bear $p-\ell -3 = 2i+1$, $i\in \mathbb{N}$, thus not contributing to the final difference. Concerning the homogeneous solution, it is apparent from Eqs.~\eqref{eq:hhom_dev} and~\eqref{eq:DeltajxL} that it bears $q=0$. As the linear metric bears $q=1$, the homogeneous solution, once inserted into the cubic source, will produce only terms with $q=1$, thus evading the condition for a term to contribute to the difference. This crucial remark that only ``even'' terms contribute makes the cubic source become local in the end, once expanded in the near zone. Therefore, our final result~\eqref{eq:DH} will itself be local, as claimed at the end of Sec.~\ref{sec:diff}.

The only quadratic interaction that does not fall within the previous investigation is the memory type one $M_{ij}\times M_{ij}$, which is needed to compute the cubic source of the 4PN tail-of-memory (see Sec.~\ref{sec:reg4PN}). This interaction involves two time-dependent moments, so that its source reads
\begin{equation}
N(\mathbf{x},t) = \hat{n}_L\,\frac{\ell_0^{q\varepsilon}}{r^{p+q\varepsilon}}\int_1^{+\infty}\!\!\dd y\,\gamma_\frac{1-d}{2}(y)\,y^s \int_1^{+\infty}\!\!\dd z\,\gamma_\frac{1-d}{2}(z)\,z^k\,F\left(t-\frac{yr}{c}\right)\,G\left(t-\frac{zr}{c}\right)\,.
\end{equation}
In this case, the PN expansion of the solution involves five types of terms: the squares ``even-even'' and ``odd-odd'', the cross products ``even-odd'' and ``odd-even'', and the homogeneous solution, which keeps the same structure as Eq.~\eqref{eq:hhom_dev}. Nevertheless, by arguments similar as for the tail interaction, only the ``even-even'' terms contribute to the final result. Note that, although the ``odd-odd'' terms could contribute as they contain terms with $p-\ell-3\in2\mathbb{N}$, it turns out that they have $q=0$. Therefore, the same argument as the one used to discard the contribution of the homogeneous solution applies. From~\eqref{eq:source_dev_even}, we see again that the result for the difference at cubic order will be purely local.

Finally, applying the formula~\eqref{eq:DhL_DH} on each terms of the (PN-expanded) cubic source, we obtain the difference between regularizations in the form of a piece $\mathcal{D}h^{\mu\nu}$ in the metric to the cubic order,
\begin{align}\label{eq:DhL_DHmunu}
\mathcal{D}h^{\mu\nu} = \sum_{\ell\in\mathbb{N}} \hat{\partial}_L\,\biggl[\frac{\tilde{k}}{r^{d-2}} \,\int_1^{+\infty} \dd y \,\gamma_{\frac{1-d}{2}}(y) \,\mathcal{D}H^{\mu\nu}\left(t-\frac{yr}{c}\right)\biggr]\,.
\end{align}
We are almost done but, while this piece clearly obeys $\Box\mathcal{D}h^{\mu\nu}=0$, it does not \textit{a priori} satisfy the harmonic coordinate condition by itself, $\partial_\nu\mathcal{D}h^{\mu\nu}\not=0$. Next, we compute from it another object, $\mathcal{D}h^{\mu\nu}_\text{gen}$, which happens to be a \textit{general} solution of the linearized Einstein field equations in vacuum (and $d$ dimensions), \textit{i.e.} which satisfies at once $\Box\mathcal{D}h^{\mu\nu}_\text{gen}=0$ and $\partial_\nu\mathcal{D}h^{\mu\nu}_\text{gen}=0$. In doing so, we are exactly following the MPM procedure~\cite{BD86, B98mult},
which can be summarized as
\begin{align}\label{eq:algH}
u^{\mu\nu}\equiv\mathcal{D}h^{\mu\nu}~\longrightarrow~ w^\mu=\partial_\nu u^{\mu\nu}~\longrightarrow~ v^{\mu\nu}=\mathcal{H}(w^\mu)~\longrightarrow~\mathcal{D}h^{\mu\nu}_\text{gen}\equiv u^{\mu\nu}+v^{\mu\nu}\,,
\end{align}
$v^{\mu\nu}$ being here a homogeneous solution such that $\partial_\nu v^{\mu\nu}= - w^\mu$, built from $w^\mu$ with the help of the ``harmonicity'' algorithm $\mathcal{H}$ given by Eqs.~(2.11)--(2.12) in~\cite{B98quad}. Since $\mathcal{D}h^{\mu\nu}_\text{gen}$ solves the linearized vacuum field equations, it can be written in a unique way as
\begin{align}\label{eq:algH} 
\mathcal{D}h^{\mu\nu}_\text{gen} = \mathcal{D}h^{\mu\nu}_\text{can} + \text{(linearized gauge transformation)}\,,
\end{align}
where the gauge transformation can be ignored as it plays no role when looking for the invariants in the metric. The physically relevant quantity is another object, $\mathcal{D}h^{\mu\nu}_\text{can}$, given as some simpler \textit{canonical} solution, which naturally takes the same form as the linearized metric~\eqref{eq:hlin_MSK}, \textit{i.e.}
\begin{subequations}\label{eq:Dhlin_MSK}
	\begin{align}
& \mathcal{D}h^{00}_\text{can}=
	- \frac{4}{c^2}\sum_{\ell \geqslant 0}\frac{(-)^\ell}{\ell !} \,\hat{\partial}_L\,\delta\widetilde{\mathcal{M}}_L\,,\\
&
\mathcal{D}h^{0i}_\text{can} =
\frac{4}{c^3}\sum_{\ell \geqslant 1}\frac{(-)^\ell}{\ell !}\,\left[\hat{\partial}_{L-1}\,\delta\widetilde{\mathcal{M}}_{iL-1}^{(1)}+\frac{\ell}{\ell+1}\hat{\partial}_L\delta\widetilde{\mathcal{S}}_{i\vert L}\right]\,,\\
&
\mathcal{D}h^{ij}_\text{can} =
- \frac{4}{c^4}\sum_{\ell \geqslant 2}\frac{(-)^\ell}{\ell !}\,\left[
\hat{\partial}_{L-2}\,\delta\widetilde{\mathcal{M}}_{ijL-2}^{(2)}+\frac{2\ell}{\ell+1}\hat{\partial}_{L-1}\delta\widetilde{\mathcal{S}}^{(1)}_{(i\vert j)L-1}+\frac{\ell-1}{\ell+1}\hat{\partial}_L\delta\widetilde{\mathcal{K}}_{ij\vert L}\right]\,,
\end{align}
\end{subequations}
where we have straightforwardly defined
\begin{equation}\label{eq:tildedDML}
\delta\widetilde{\mathcal{M}}_L(r,t) \equiv \frac{\tilde{k}}{r^{d-2}}\int_1^{+\infty}\!\!\dd y\,\gamma_\frac{1-d}{2}(y)\,\mathcal{D}M_L\left(t-\frac{yr}{c}\right)\,,
\end{equation}
and where $\mathcal{D}M_L$ and $\mathcal{D}S_{i\vert L}$ denote the corresponding corrections to the mass and current canonical moments $M_L$ and $S_{i\vert L}$, together with those of the additional moment $K_{ij\vert L}$, which does not exist in 3 dimensions. In conclusion, when written into the form~\eqref{eq:Dhlin_MSK}, our result is directly formatted to read off the non-linear corrections to the canonical moments. From the detailed calculations of the cubic interactions (in particular $M\times M_{ij}\times M_{ij}$) outlined below, and the comparison with~\eqref{eq:Dhlin_MSK}--\eqref{eq:tildedDML}, we will find $\delta\widetilde{\mathcal{M}}_{ij}(r,t)$ and, thus, the looked-for correction to the canonical quadrupole moment $\mathcal{D}M_{ij}(t)$. This result will play a crucial role in the definition of the ``\textit{renormalized mass quadrupole}'' in Sec.~\ref{sec:reg4PN}.

Finally, we have used two methods to extract the correction to the canonical moment $\delta\widetilde{\mathcal{M}}_{ij}$ from the metric. One method consists of considering the leading order $1/r^{d-2}$ term in the metric when $r\to+\infty$ and performing a suitable angular integration (in $d$ dimensions) to obtain $\delta\widetilde{\mathcal{M}}_{ij}^{(2)}$, then $\delta\widetilde{\mathcal{M}}_{ij}$.
The second method was to extract it directly from the gauge invariant components of the linearized Riemann tensor
\begin{equation}
\hspace{-0.2cm}\mathcal{R}_{0i0j} = \frac{G}{2}\left(\partial_{ij}\mathcal{D}h^{00}_\text{can}+ \frac{2}{c}\,\partial_t\partial^{(i}\mathcal{D}h^{j)0}_\text{can} + \frac{1}{c^2}\,\partial_t^2\mathcal{D}h^{ij}_\text{can} + \frac{\partial_{ij}\mathcal{D}h_\text{can}}{d-1}-\frac{1}{c^2} \frac{\delta_{ij}}{d-1}\,\partial_t^2\mathcal{D}h_\text{can}\right)\,,
\end{equation}
where $\mathcal{D}h_\text{can} \equiv -\mathcal{D}h^{00}_\text{can}+\mathcal{D}h^{ii}_\text{can}$. Both methods perfectly agree with each other.

\section{Regularization of non-linear interactions at 3PN order}
\label{sec:reg3PN}

When computing the radiative mass quadrupole moment at 3PN order~\cite{BDEI05dr}, a Hadamard regularization scheme for the IR sector has been used for the source moment and the non-linear interactions have been computed in ordinary 3 dimensions. However, the crucial need for a dimensional regularization scheme in the 4PN equations of motion~\cite{BBBFMc} implies that one should trade the Hadamard regularization for dimensional regularization when computing the mass quadrupole at the 4PN order. This has been achieved in the companion paper~\cite{DDR_source}. It was found that, when using a dimensional regularization scheme, the source quadrupole moment receives corrections including poles $\propto\varepsilon^{-1}$ already starting at the 3PN order.

To describe the poles, it is convenient to ``dress'' them into the particular combination
\begin{equation}\label{eq:pole_habille}
\Pi_\varepsilon \equiv -\frac{1}{2\varepsilon}+\ln\left(\frac{r_0\sqrt{\bar{q}}}{\ell_0}\right)\,,
\quad
\text{with}\quad
\bar{q} \equiv 4\pi \de^{\gamma_E}\,.
\end{equation}
Here $r_0$ is the scale associated to the Partie Finie and $\ell_0$ the one associated to the dimensional regularization. Now, the detailed calculations in~\cite{DDR_source} conclude that, at the 3PN order, the dimensional regularization corrections can be recast as
\begin{equation}\label{eq:poles_Iij_3PN}
\mathcal{D}I_{ij}^\text{3PN} = 
\frac{214}{105}\,\frac{G^2M^2}{c^6}\left(\Pi_\varepsilon+\frac{246\,299}{44\,940}\right)\,I_{ij}^{(2)}-\frac{428}{105}\,\frac{G^2M}{c^6}\left(\Pi_\varepsilon+\frac{252\,599}{44\,940}\right)\,P_{\langle i}P_{j\rangle}\,,
\end{equation}
where $I_{ij}$ is the source quadrupole moment and $P_i$ is the conserved linear momentum, \textit{i.e.} $P_i=I^{(1)}_i$, with $I_i$ being the linearly varying mass dipole. Note that this formula is valid in an arbitrary frame, not necessarily the center-of-mass (CoM) frame for which $P_i=0$. Obviously, $I_{ij}$ and $P_i$ are merely Newtonian with the 3PN accuracy of~\eqref{eq:poles_Iij_3PN} but will contain crucial 1PN corrections when this equation is applied to 4PN order. 

We recognize $\beta_I = -\frac{214}{105}$ in the coefficient of the first term in~\eqref{eq:poles_Iij_3PN}. This coefficient is well known from ``tail-of-tail'' calculations~\cite{B98tail}. It can be interpreted as the beta function coefficient associated with the logarithmic renormalization of the mass quadrupole moment~\cite{GRoss10}. 

The 3PN correction to the source quadrupole~\eqref{eq:poles_Iij_3PN} bears a pole, which looks dreadful. Fortunately, the source quadrupole is not a physical quantity, contrary to the radiative quadrupole defined at infinity. We thus have to ensure that the correction~\eqref{eq:poles_Iij_3PN} is exactly compensated by the regularization of the 3PN non-linear interactions in the radiative moment. Otherwise, it would show that the formalism used previously at 3PN order is inconsistent with dimensional regularization.

A trivial dimensional analysis shows that only two interactions can arise at 3PN order: (i) the tail-of-tail $M\times M\times M_{ij}$, where $M$ is the constant ADM mass and $M_{ij}$ the canonical moment (which agrees with the source moment to 2PN order), and (ii) the interaction $M\times M_i\times M_i$ where $M_i$ is the mass dipole moment, varying linearly with time and vanishing in the CoM frame. As we have proved in Sec.~\ref{sec:diff} [see~\eqref{eq:DH}], the regularization of the non-linear interactions linking the canonical and radiative quadrupoles yields purely local corrections. Moreover, since in general relativity the propagation of GW is described entirely by the two types of multipole moments $M_L$ and $S_{i\vert L}$ (called the canonical moments in our terminology) we have to express the non-linear terms as corrections in the canonical moments, as discussed in Sec.~\ref{sec:MPM} [see~\eqref{eq:Dhlin_MSK}].

Implementing the non-linear iteration of the requested multipole interactions, as described in Sec.~\ref{sec:MPM}, we computed the 3PN correction to the canonical quadrupole:
\begin{equation}\label{eq:poles_Mij_3PN}
\mathcal{D}M_{ij}^\text{3PN} = -\frac{214}{105}\,\frac{G^2M^2}{c^6}\left(\Pi_\varepsilon+\frac{246\,299}{44\,940}\right)\,M_{ij}^{(2)}
+\frac{428}{105}\,\frac{G^2M}{c^6}\left(\Pi_\varepsilon+\frac{252\,599}{44\,940}\right)\,M^{(1)}_{\langle i}M^{(1)}_{j\rangle}\,.
\end{equation}
As the canonical and source moments are equivalent at Newtonian order, we see that the contribution of the non-linear interactions exactly cancels the correction of the source quadrupole given by Eq.~\eqref{eq:poles_Iij_3PN}. Again, this result is valid in a general frame, not necessarily in the CoM frame, and it is true not only for the pole part, but also for the finite part following the pole. 

We conclude that the corrections due to dimensional IR regularization exactly vanish in the radiative quadrupole to 3PN order; so, using Hadamard or dimensional regularization for the IR is equivalent at that order. \textit{A priori}, the same cancellation is not expected to happen at 4PN order, because we found for the conservative equations of motion that the dimensional and Hadamard regularization schemes are no more equivalent~\cite{BBBFMc}. Correcting the quadrupole moment to account for dimensional regularization at 4PN order is the main motivation for this work.

Two other radiative moments are very important as they have to be computed to 3PN order in order to control the 4PN flux: these are the mass octupole moment, obtained in~\cite{FBI15}, and the current quadrupole moment, computed in~\cite{HFB_courant}. We have to verify the equivalence of both regularization schemes for these 3PN moments too. Following the methods described in~\cite{DDR_source}, we have computed the correction due to the change in regularization schemes for both the source mass octupole $I_{ijk}$ and the (dual of the) current quadrupole $J_{i\vert jk}$, with results
\begin{subequations}\label{eq:DIJ}
\begin{align}
\mathcal{D}I_{ijk}^\text{3PN} &= \frac{26}{21}\frac{G^2}{c^6}\left(\Pi_\varepsilon+\frac{9281}{2730}\right)M^2I_{ijk}^{(2)}
+\frac{92}{35}\frac{G^2}{c^6}\left(\Pi_\varepsilon-\frac{161\,597}{28\,980}\right)I_{\langle i}P_jP_{k\rangle}\nonumber\\
& -\frac{52}{7}\frac{G^2}{c^6}\left(\Pi_\varepsilon+\frac{14\,387}{4\,095}\right)M I_{\langle ij}^{(1)}P_{k\rangle}
+\frac{12}{5}\frac{G^2}{c^6}\left(\Pi_\varepsilon+\frac{63\,421}{7560}\right)M I_{\langle ij}^{(2)}I_{k\rangle}\,,\\
\mathcal{D} J_{i\vert jk}^\text{3PN} &= \frac{214}{105}\frac{G^2}{c^6}\left(\Pi_\varepsilon+\frac{54\,989}{44\,940}\right)M^2J_{i\vert jk}^{(2)} \nonumber\\
& +\frac{856}{105}\frac{G^2}{c^6}\left(\Pi_\varepsilon+\frac{229\,289}{44\,940}\right)M I_{\lbrace i\vert j}^{(2)}P_{k\rbrace}
-\frac{106}{7}\frac{G^2}{c^6}\,M I_{\lbrace i\vert j}^{(3)}I_{k\rbrace}\,.
\end{align}
\end{subequations}
Here $\lbrace i\vert jk\rbrace \equiv \underset{ij}{\mathcal{A}}\,\underset{ijk}{\text{TF}}\,\underset{jk}{\text{STF}}$ denotes the appropriate symmetries of the $d$-dimensional current quadrupole derived in~\cite{HFB_courant} [see Eq.~(2.36) there]. In Eqs.~\eqref{eq:DIJ}, we again recognize the usual beta function coefficients $\beta_O = -26/21$ and $\beta_J = -214/105$ ($=\beta_I$)\footnote{These coefficients are known for both mass and current tail-of-tail interactions up to any multipolar order~\cite{BD88,AFS21}.} in front of the tail-of-tail contribution to the octupole and current quadrupole. On the other hand, using the formulas of section~\ref{sec:diff}, we derived the corrections to those canonical moments as
\begin{subequations}\label{eq:DMS}
\begin{align}
\mathcal{D}M_{ijk}^\text{3PN} &=
-\frac{26}{21}\frac{G^2}{c^6}\left(\Pi_\varepsilon+\frac{9281}{2730}\right)M^2M_{ijk}^{(2)}
-\frac{92}{35}\frac{G^2}{c^6}\left(\Pi_\varepsilon-\frac{161\,597}{28\,980}\right)M_{\langle i}M^{(1)}_jM^{(1)}_{k\rangle}\nonumber\\
& +\frac{52}{7}\frac{G^2}{c^6}\left(\Pi_\varepsilon+\frac{14\,387}{4\,095}\right)M M_{\langle ij}^{(1)}M^{(1)}_{k\rangle}
-\frac{12}{5}\frac{G^2}{c^6}\left(\Pi_\varepsilon+\frac{63\,421}{7560}\right)M M_{\langle ij}^{(2)}M_{k\rangle}\,,\\
\mathcal{D}S_{i\vert jk}^\text{3PN} &= 
-\frac{214}{105}\frac{G^2}{c^6}\left(\Pi_\varepsilon+\frac{54\,989}{44\,940}\right)M^2S_{i\vert jk}^{(2)}\nonumber\\
& -\frac{856}{105}\frac{G^2}{c^6}\left(\Pi_\varepsilon+\frac{229\,289}{44\,940}\right)M M_{\lbrace i\vert j}^{(2)}M^{(1)}_{k\rbrace}
+\frac{106}{7}\frac{G^2}{c^6}\,M M_{\lbrace i\vert j}^{(3)}M_{k\rbrace}\,.
\end{align}
\end{subequations}
Similarly to the case of the 3PN mass quadrupole, we find that the proper regularization of the non-linear interactions exactly compensates the correction due to the dimensional regularization of the source mass octupole and current quadrupole. As clear from~\eqref{eq:DIJ} and~\eqref{eq:DMS}, this is proven not only for the pole parts but also for the finite terms beyond the poles. In other words, for all these 3PN moments, the corrections due to dimensional regularization exactly vanish in the radiative moments, which are the physically relevant quantities. In conclusion, the present investigations have confirmed the previous calculations of multipole moments at the 3PN order presented in Refs.~\cite{BI04mult, BDEI05dr, FBI15, HFB_courant}.

\section{The renormalized mass quadrupole moment at 4PN order}
\label{sec:renorm4PN}

\subsection{Regularization of non-linear interactions at 4PN order}
\label{sec:reg4PN}

Having proved that Hadamard and dimensional (IR) regularizations are equivalent at the 3PN order, we now turn to the main goal of this work, namely the IR regularization of the radiative mass quadrupole up to 4PN.

In addition to the 3PN interaction~\eqref{eq:poles_Mij_3PN} which has now to be considered up to 4PN, five additional interactions can show up: the ``tail-of-memory'' $M\times M_{ij}\times M_{ij}$, the interaction $M\times M_{ij}\times S_{i\vert j}$ involving the constant angular momentum $S_{i\vert j}$, and three interactions $M\times M_i\times M_{ijk}$, $M_i\times M_i\times M_{ij}$ and $M\times M_i\times S_{i\vert jk}$ involving the mass dipole $M_i$. As discussed earlier, the difference between the regularizations of the non-linear interactions are local and can be expressed as corrections to the canonical moments $M_L$ and $S_{i\vert L}$.

For the sake of simplicity, we will work in the CoM frame, where $M_i$ vanishes. Indeed, the main application of the present work for PN templates can be performed in such frame. Only three interactions survive, yielding the following corrections in the canonical moment: 
\begin{align}\label{eq:poles_Mij_4PN}
\mathcal{D}M_{ij}^\text{CoM, 4PN} =\, &
-\frac{214}{105}\,\frac{G^2M^2}{c^6}\left(\Pi_\varepsilon+\frac{246\,299}{44\,940}\right)\,M_{ij}^{(2)}\\
& +\frac{G^2M}{c^8}\Biggl[
\frac{12}{7}\left(\Pi_\varepsilon+\frac{3581}{7560}\right)\,M_{a\langle i}^{(2)}\,M_{j\rangle a}^{(2)}
-\frac{24}{7}\left(\Pi_\varepsilon-\frac{338}{2835}\right)\,M_{a\langle i}^{(1)}\,M_{j\rangle a}^{(3)}\nonumber\\
& \qquad\qquad
-\frac{4}{7}\left(\Pi_\varepsilon-\frac{1447}{216}\right)\,M_{a\langle i}\,M_{j\rangle a}^{(4)}
+\frac{4}{3}\left(\Pi_\varepsilon+\frac{11\,243}{7560}\right)M_{a\langle i}^{(3)}S_{j\rangle \vert a}\Biggl]\,,\nonumber
\end{align}
where we recall that the pole has been ``dressed-up'' according to~\eqref{eq:pole_habille}.

As we have discussed in Sec.~\ref{sec:reg3PN} the tail-of-tail contribution given by the first line of~\eqref{eq:poles_Mij_4PN} exactly cancels the correction due to dimensional regularization in the source moment at the 3PN order. Note that this keeps being true up to the 4PN order because the canonical and source moments are equivalent at the 1PN order. Now, we discover that the coefficients of the remaining poles entering the non-linear interactions in the purely 4PN terms of~\eqref{eq:poles_Mij_4PN} exactly cancel those coming from the IR regularization of the source mass quadrupole, given by Eq.~(4.4) of the companion paper~\cite{DDR_source}.

Therefore, we arrive at the main result of this paper, namely that the radiative mass quadrupole moment is free of poles up to the 4PN level. This important feature (but expected for a physical quantity) naturally leads us to define a ``\textit{renormalized mass quadrupole}'' as the sum of the properly regularized source mass quadrupole, which was computed in~\cite{MHLMFB20, DDR_source}, and augmented by the contributions from the dimensional regularization of the cubic non-linear interactions, as displayed in Eqs.~\eqref{eq:poles_Mij_4PN}, \textit{including} the finite corrections beyond the poles explicitly found in~\eqref{eq:poles_Mij_4PN}. The interest of such quantity is that it is finite when $\varepsilon\to 0$, so that it can be considered and manipulated in ordinary three dimensions. This notably allows us to express it in the CoM frame and, then, for quasi-circular orbits using the usual three-dimensional reduction procedure. 

Recalling the work conducted in~\cite{DDR_source}, the renormalized quadrupole is thus defined as
\begin{equation}\label{eq:Iijrenorm}
I^\text{renorm}_{ij} \equiv I_{ij} + \mathcal{D}M_{ij}
= I_{ij}^\text{Had} + I_{ij}^\text{non-loc} + \mathcal{D}I_{ij} +\delta_\chi I_{ij} + \mathcal{D}M_{ij}\,,
\end{equation}
where $I_{ij}^\text{Had}$ denotes the result of~\cite{MHLMFB20}, \textit{i.e.} the local source quadrupole computed with a UV dimensional regularization and an IR Hadamard one; $I_{ij}^\text{non-loc}$ is the non-local effect of the \textit{source} quadrupole, coming from the 4PN tail effect in the conservative sector (equations of motion and Fokker Lagrangian), given in Eq.~(2.14) of~\cite{DDR_source}; $\mathcal{D}I_{ij}$ is the correction due to dimensional regularization in the source quadrupole ($\mathcal{D}I_{ij}$ is itself the sum of four contributions in Eq.~(3.1) of~\cite{DDR_source}); $\delta_\chi I_{ij}$ is the effect of the local part of the IR shift applied in the conservative sector~\cite{DDR_source, BBBFMc}; and, finally, $\mathcal{D}M_{ij}$ is the correction due to dimensional regularization of the non-linear interactions, computed in the present work. 

The non-local tail term in the source quadrupole found in~\cite{DDR_source} can be restated as
\begin{align}\label{eq:nonlocdecomp}
I_{ij}^\text{non-loc} = I_{ij}^\text{inst} + \hat{I}_{ij}^\text{non-loc}\,.
\end{align}
The first term corresponds to instantaneous (non-tail) contributions given by
\begin{align}\label{eq:tailinst}
\hat{I}_{ij}^\text{inst} = \frac{24}{7}\frac{G^2M}{c^8}\left(\Pi_\varepsilon+\frac{74}{105}\right)\,I_{a\langle i}\,I_{j\rangle a}^{(4)}\,,
\end{align}
and which combine with~\eqref{eq:poles_Mij_4PN} and the other terms in~\eqref{eq:Iijrenorm} to cancel the pole. The second term is, properly speaking, the purely non-local tail contribution which reads as
\begin{align}\label{eq:puretail}
\hat{I}_{ij}^\text{non-loc} = \frac{24}{7} \frac{G^2M}{c^{8}} \,I_{k\langle i}(t)\int_0^{+\infty}\dd\tau\, \ln\left(\frac{c\tau}{2r_0}
\right)\,I^{(5)}_{j\rangle k}(t-\tau)\,.
\end{align}
Such contribution will naturally combine with the non-localities induced by the non-linear interactions in the radiative quadrupole. The non-local tail term contains both conservative and dissipative effects,
\begin{equation}
\hat{I}_{ij}^\text{non-loc} = \hat{I}_{ij}^\text{non-loc}\bigg|_\text{cons} + \hat{I}_{ij}^\text{non-loc}\bigg|_\text{diss}\,,
\end{equation}
where the conservative sector is given by the time-symmetrized integral
\begin{align}\label{eq:puretailcons}
\hat{I}_{ij}^\text{non-loc}\bigg|_\text{cons} = \frac{12}{7} \frac{G^2M}{c^{8}} \,I_{k\langle i} \int_0^{+\infty}\dd\tau\, \ln\left(\frac{c\tau}{2r_0}
\right)\left[I^{(5)}_{j\rangle k}(t-\tau) - I^{(5)}_{j\rangle k}(t+\tau)\right]\,,
\end{align}
or can also be expressed in the form of the Hadamard partie finie (Pf) integral
\begin{align}\label{eq:puretailconsHad}
\hat{I}_{ij}^\text{non-loc}\bigg|_\text{cons} = \frac{12}{7} \frac{G^2M}{c^{8}} \,I_{k\langle i}\,\text{Pf}_{2r_0/c} \int_{-\infty}^{+\infty}\frac{\dd t'}{\vert t-t'\vert} \,I^{(4)}_{j\rangle k}(t')\,,
\end{align}
with associated time scale $2r_0/c$. 
The dissipative effect is naturally encoded in the complementary time-antisymmetrized integral
\begin{align}\label{eq:puretaildiss}
\hat{I}_{ij}^\text{non-loc}\bigg|_\text{diss} = \frac{12}{7} \frac{G^2M}{c^{8}} \,I_{k\langle i} \int_0^{+\infty}\dd\tau\, \ln\left(\frac{c\tau}{2r_0}
\right)\left[I^{(5)}_{j\rangle k}(t-\tau) + I^{(5)}_{j\rangle k}(t+\tau)\right]\,.
\end{align}

Note that the notion of renormalized mass quadrupole applies to the source type moment, hence we appropriately called it $I^\text{renorm}_{ij}$ and not $M^\text{renorm}_{ij}$. In order to obtain the physical radiative moment, we still have to add to the renormalized source moment the corrections coming from (i) the relation between the canonical quadrupole on the one hand, and the source and gauge moments on the other hand (see~\cite{B98mult} for discussion), and (ii) the non-linear tails, tails-of-tails, tails-of-memory, \textit{etc.} multipole interactions. However, the interest of having defined such renormalized source quadrupole moment $I^\text{renorm}_{ij}$ is that the latter corrections (i) and (ii) can be computed in ordinary three dimensions using the standard MPM algorithm, as all dimensional regularization contributions have already been included. Many of the corrections (i)-(ii) are already known (see notably~\cite{BFIS08, FMBI12}), although not all of them at the 4PN order. Notably, the involved cubic tail-of-memory terms $M\times M_{ij}\times M_{ij}$ will still have to be computed in 3 dimensions, which is left for future work.

\subsection{Center-of-mass frame and quasi-circular orbits}
\label{sec:circ}

As, by construction, the renormalized quadrupole is a three-dimensional quantity, one can safely reduce it to the CoM frame and, then, specialize it to the case of quasi-circular orbits, using usual three-dimensional reduction formulas. Unfortunately, the complete quadrupole is too long to be presented, even when expressed in the CoM frame. Therefore, we will only present the difference between the purely local renormalized quadrupole and the Hadamard one $\Delta I_{ij} \equiv  I^\text{renorm}_{ij}-\hat{I}_{ij}^\text{non-loc}-I^\text{Had}_{ij}$ in the CoM frame. 

We denote by $r=\vert\bm{y}_1-\bm{y}_2\vert$ the radial separation in harmonic coordinates, $\bm{x}=\bm{y}_1-\bm{y}_2$ the relative distance, and $\bm{v}=\bm{v}_1-\bm{v}_2$ the relative velocity. We recall also the dimensionless 1PN parameter $\gamma$ defined as $\gamma = \frac{G m}{r c^2}$ with $m = m_1+m_2$ the total mass and $\nu = m_1 m_2/m^2$ the symmetric mass ratio. The difference $\Delta I_{ij}$ is rather compact, when written in the CoM frame. It reads
\begin{align}\label{eq:DeltaIijCoM}
\Delta I_{ij}^\text{CoM} =& \frac{G^2m^3\nu}{c^8}\Biggl\{
\frac{G^2m^2}{r^2}\left[\frac{1072}{105}-\frac{571826}{14175}\nu\right]n^{\langle i}n^{j\rangle}
+\left[\frac{7516}{735}-\frac{123944}{6615}\nu\right]v^2\,v^{\langle i}v^{j\rangle}\\
& 
\quad +\frac{G m}{r}\biggl(\left[\frac{3004}{147}+\frac{1800322}{11025}\nu\right](n.v)^2\,n^{\langle i}n^{j\rangle}
+\left[-\frac{30004}{2205}-\frac{9985783}{99225}\nu\right]v^2\,n^{\langle i}n^{j\rangle}\nonumber\\
& 
\qquad\qquad+\left[-\frac{7516}{245}-\frac{2437018}{19845}\nu\right](n.v)\,n^{\langle i}v^{j\rangle}
+\left[-\frac{15056}{2205}+\frac{161096}{6615}\nu\right]v^{\langle i}v^{j\rangle}\biggr)\Biggr\}\,.\nonumber
\end{align}
It is fully apparent that the dimensional regularization brings in a finite contribution to the quadrupole moment at the 4PN order. In previous work on 4PN equations of motion~\cite{BBBFMc, MBBF17}, adding the analogue of~\eqref{eq:DeltaIijCoM} permitted resolving the long-standing problem of regularization ambiguities.

Finally, we give the complete end result for the renormalized quadrupole on quasi-circular orbits. We write it in the form\footnote{Note that, after publishing the Hadamard source quadrupole in~\cite{MHLMFB20}, we have spotted an error in the $d$-dimensional computation of the value of the potential $\hat{R}_i$ at 1PN order, when evaluated in $\bm{y}_{1,2}$. This error induces a small change in the value of the coefficients, namely $\delta A = -4\gamma^4\nu^2/63$ and $\delta B = 4\gamma^3\nu^2/63$, which has been taken into account here.}
\begin{equation}\label{eq:Iijcirc}
I_{ij}^\text{renorm} = m\nu\, \left(
A \, x_{\langle i}x_{j \rangle}
+B \, \frac{r^2}{c^2}v_{\langle i}v_{j \rangle}
+ \frac{G^2 m^2\nu}{c^5r}\,C\,x_{\langle i}v_{j \rangle}\right) + \mathcal{O}\left(\frac{1}{c^{9}}\right)\,,
\end{equation}
where the coefficients read
\begin{subequations}\label{eq:Iijrenorm_AB}
	\begin{align}
	A &= 1
	+ \gamma \biggl(- \frac{1}{42}
	-  \frac{13}{14} \nu \biggr)
	+ \gamma^2 \biggl(- \frac{461}{1512}
	-  \frac{18395}{1512} \nu
	-  \frac{241}{1512} \nu^2\biggr)
	\nonumber\\
	& \quad + \gamma^3 \biggl(\frac{395899}{13200}
	-  \frac{428}{105} \ln\biggl(\frac{r}{r_{0}{}} \biggr)
	+ \biggl[\frac{3304319}{166320}
	-  \frac{44}{3} \ln\biggl(\frac{r}{r'_{0}}\biggr) \biggr]\nu
	+ \frac{162539}{16632} \nu^2 + \frac{2351}{33264} \nu^3
	\biggr)
	\nonumber\\
	&  \quad + \gamma^4 \biggl (- \frac{1067041075909}{12713500800}
	+ \frac{31886}{2205} \ln\biggl(\frac{r}{r_{0}{}} \biggr)
	+ \biggl[-\frac{85244498897}{470870400}
	-  \frac{2783}{1792} \pi^2-\frac{64}{7}\ln\left(16\gamma \de^{2\gamma_\text{E}}\right) \nonumber\\
	& \qquad \quad -  \frac{10886}{735} \ln\biggl(\frac{r}{r_{0}{}} \biggr)
	+ \frac{8495}{63} \ln\biggl(\frac{r}{r'_{0}} \biggr)\biggr] \nu
	+ \biggl[\frac{171906563}{4484480} + \frac{44909}{2688} \pi^2-  \frac{4897}{21}
	\ln\biggl(\frac{r}{r'_{0}} \biggr)\biggr]\nu^2\nonumber\\
	& \qquad \quad - \frac{22063949}{5189184} \nu^3 + \frac{71131}{314496} \nu^4
	\biggl)\,, \\
	B &=\frac{11}{21}
	-  \frac{11}{7} \nu
	+ \gamma \biggl(\frac{1607}{378}
	-  \frac{1681}{378} \nu
	+ \frac{229}{378} \nu^2\biggr) \nonumber\\
	& \quad + \gamma^2 \biggl(- \frac{357761}{19800}
	+ \frac{428}{105} \ln\biggl(\frac{r}{r_{0}{}} \biggr)
	-  \frac{92339}{5544} \nu
	+ \frac{35759}{924} \nu^2
	+ \frac{457}{5544} \nu^3 \biggr)  \nonumber\\
	& \quad + \gamma^3 \biggl(\frac{23006898527}{1589187600} -  \frac{4922}{2205}
	\ln\biggl(\frac{r}{r_{0}{}} \biggr) + \biggl[\frac{8431514969}{529729200}
	+ \frac{143}{192} \pi^2-\frac{32}{7}\ln\left(16\gamma \de^{2\gamma_\text{E}}\right) \nonumber\\
	& \qquad \quad
	-  \frac{1266}{49} \ln\biggl(\frac{r}{r_{0}{}} \biggr)
	- \frac{968}{63} \ln\biggl(\frac{r}{r'_{0}} \biggr)\biggr] \nu  \nonumber\\
	& \qquad \quad + 
	\biggl[\frac{351838141}{5045040} 
	-  \frac{41}{24} \pi^2
	+ \frac{968}{21} \ln\biggl(\frac{r}{r'_{0}} \biggr)\biggr] \nu^2
	-  \frac{1774615}{81081} \nu^3
	-  \frac{3053}{432432} \nu^4 \biggl)\,, \\
	C &= \frac{48}{7} + \gamma \left(-\frac{4096}{315} - \frac{24512}{945}\nu \right)-\frac{32}{7}\pi\,\gamma^{3/2}\,.
	\end{align}
\end{subequations}
The term $C$ corresponds to time-odd contributions. The 2.5PN and 3.5PN terms were computed in~\cite{FMBI12}, whereas the new 4PN term $-\frac{32}{7}\pi\,\gamma^{3/2}$ comes from the dissipative contribution of the non-local term~\eqref{eq:puretaildiss}. As for the conservative contribution of the tail term~\eqref{eq:puretailconsHad}, it yields the logarithmic dependences $\ln(16\gamma \de^{2\gamma_\text{E}})$ in the time-even coefficients $A$ and $B$. 

Finally, the two constants $r_0$ and $r_0'$ entering the coefficients~\eqref{eq:Iijrenorm_AB} are associated with the Partie Finie and the dimensional UV regularization schemes, respectively (see~\cite{MHLMFB20}). As unphysical scales, they are expected to be exactly compensated in the radiative moment, by virtue of non-linear effects and application of the time derivatives. This will be checked in future work.

\acknowledgments

We thank Gabriel Luz Almeida, Laura Bernard, Stefano Foffa, Sylvain Marsat, Rafael Porto and Riccardo Sturani for interesting discussions. 
F.L. received funding from the European Research Council (ERC) under the European Union’s Horizon 2020 research and innovation programme (grant agreement No 817791).

\appendix

\section{Check with the $d=3$ limit of the retarded integral}
\label{app:limit3d}
 
In three dimensions, the function $\gamma_{\frac{1-d}{2}}(z)$ defined by~\eqref{eq:gamma} takes the distributional form $\gamma_{-1}(z) = \delta(z-1)$, where $\delta(z-1)$ means the Dirac function at point $z=1$. More generally, we have for any $\ell\in\mathbb{N}$ (see App. C of~\cite{BBBFMc})
\begin{align}\label{eq:gammadistr}
\gamma_{-1-\ell}(z) = \frac{1}{(2\ell-1)!!}\sum_{i=0}^{\ell}
\beta_i^\ell\,\delta^{(i)}(z-1) \,,
\end{align}
where $\delta^{(i)}$ denotes the $i$-th derivative of the Dirac function. The numerical coefficients $\beta_i^\ell \equiv 2^{i-\ell}\frac{(2\ell-i)!}{i! (\ell-i)!}$ are such that the multipolar retarded or advanced homogeneous wave in 3 dimensions reads
\begin{align}\label{eq:multhom}
\hat{\partial}_L\!\left(\frac{F(t\mp r)}{r}\right) &= (-)^\ell \frac{\hat{n}_L}{r^{\ell+1}}\sum_{i=0}^\ell \beta_i^\ell \,(\pm r)^i \,F^{(i)}(t\mp r)\,,
\end{align}

We compute the 3-dimensional limit of our final $d$-dimensional result~\eqref{eq:hcomplet}. In this limit the two integrations over variables $y$ and $z$ can be effected explicitly thanks to Eq.~\eqref{eq:gammadistr}, while the integration over $\tau$ follows from the limit~\eqref{eq:limiteps0} which just amounts to replacing the source term $\widehat{N}_\varepsilon$ by $N$. Thus, we obtain in 3 dimensions
\begin{align}\label{eq:h3d0}
&h_L = \frac{(-)^\ell}{2}\,\hat{n}_L \sum_{i=0}^{\ell}\sum_{j=0}^{\ell}
\beta_i^\ell\,\beta_j^\ell \,r^{-1-\ell+j}\nonumber\\&\qquad\times\Biggl\{\int_0^{+\infty}\!\dd {r'} \,{r'}^{B-\ell+1+i}N^{(-2\ell-1+i+j)}({r'}, t - r - {r'})\nonumber\\&\qquad\qquad - (-)^j\int_r^{+\infty}\!\dd {r'} \,{r'}^{B-\ell+1+i}N^{(-2\ell-1+i+j)}({r'}, t + r - {r'})\nonumber\\&\qquad\qquad - (-)^i \int_0^r\!\dd {r'} \,{r'}^{B-\ell+1+i}N^{(-2\ell-1+i+j)}({r'}, t - r + {r'}) \Biggr\}\,.
\end{align}
After inserting the well-known formula for the multiple time anti-derivative (valid for the source term becoming zero when ${t'}\to-\infty$)
\begin{align}\label{eq:antideriv}
N^{(-2\ell-1+i+j)}({r'}, t')=\int_{-\infty}^{t'-r'}\dd s\,\frac{(t'-r'-s)^{2\ell-i-j}}{(2\ell-i-j)!}N({r'},s+{r'}) \,,
\end{align}
and inverting the summations over ${r'}$ and $s$, we can re-express the result~\eqref{eq:h3d0} into the form
\begin{align}\label{eq:h3d1}
&h_L = \int_{-\infty}^{t-r}\dd s\,\Biggl\{\int_0^{\frac{t-r-s}{2}}\!\dd {r'} \,{r'}^{B-\ell+1}N({r'},s+{r'})\,\mathcal{A}_L\\&\qquad\qquad\qquad - \int_r^{\frac{t+r-s}{2}}\!\dd {r'} \,{r'}^{B-\ell+1}N({r'},s+{r'})\,\mathcal{B}_L - \int_0^r\!\dd {r'} \,{r'}^{B-\ell+1}N({r'},s+{r'})\,\mathcal{C}_L \Biggr\}\,.\nonumber
\end{align}
Here, we have posed as intermediate notation
\begin{subequations}\label{eq:ABCL}
\begin{align}
\mathcal{A}_L &= \frac{(-)^\ell}{2}\frac{\hat{n}_L}{r^{1+\ell}} \sum_{i=0}^{\ell}\sum_{j=0}^{\ell}
\beta_i^\ell\,\beta_j^\ell \,r^{j}{r'}^i\frac{(t-r-s-2{r'})^{2\ell-i-j}}{(2\ell-i-j)!}\,,\\
\mathcal{B}_L &= \frac{(-)^\ell}{2}\frac{\hat{n}_L}{r^{1+\ell}} \sum_{i=0}^{\ell}\sum_{j=0}^{\ell}
\beta_i^\ell\,\beta_j^\ell \,(-r)^{j}{r'}^i\frac{(t+r-s-2{r'})^{2\ell-i-j}}{(2\ell-i-j)!}\,,\\
\mathcal{C}_L &= \frac{(-)^\ell}{2}\frac{\hat{n}_L}{r^{1+\ell}} \sum_{i=0}^{\ell}\sum_{j=0}^{\ell}
\beta_i^\ell\,\beta_j^\ell \,r^{j}(-{r'})^i\frac{(t-r-s)^{2\ell-i-j}}{(2\ell-i-j)!}\,.
\end{align}
\end{subequations}
All the point is now to rewrite these quantities into a more convenient form. For this we dispose of the following formula, valid for any $\rho$ and ${r'}$,
\begin{align}\label{eq:formule1}
\sum_{i=0}^\ell \beta_i^\ell \,\frac{{r'}^i(2\rho-2{r'})^{2\ell-i-j}}{(2\ell-i-j)!} = \frac{2^{\ell-j}}{\ell!}\left(\frac{\dd}{\dd\rho}\right)^j\Bigl[\rho^\ell(\rho-{r'})^\ell\Bigr]\,,
\end{align}
together with the fact that the coefficients $\beta_i^\ell$ also enter the formula~\eqref{eq:multhom}, which can alternatively be further written as
\begin{align}\label{eq:multhomalt}
\left(\frac{1}{r}\frac{\partial}{\partial r}\right)^\ell \left[\frac{F(t\mp r)}{r}\right] = \frac{(-)^\ell}{r^{2\ell+1}}\sum_{i=0}^\ell \beta_i^\ell \,(\pm r)^i \,F^{(i)}(t\mp r)\,.
\end{align}
These two facts permit proving that the three terms defined in~\eqref{eq:ABCL} are actually all identical:
\begin{equation}\label{eq:ABCequal} \mathcal{A}_L=\mathcal{B}_L=\mathcal{C}_L\quad\text{(for any $t$, $s$, $r$, ${r'}$)}\,,
\end{equation} 
so that we may glue together the three integrals in~\eqref{eq:h3d1} and arrive at
\begin{align}\label{eq:h3d2}
&h_L = \int_{-\infty}^{t-r}\dd s\,\Biggl\{\int_0^{\frac{t-r-s}{2}}\!\dd {r'} \,{r'}^{B-\ell+1}N({r'},s+{r'})\,\mathcal{A}_L - \int_0^{\frac{t+r-s}{2}}\!\dd {r'} \,{r'}^{B-\ell+1}N({r'},s+{r'})\,\mathcal{B}_L \Biggr\}\,,
\end{align}
together with the following summed-up expressions 
\begin{subequations}\label{eq:ABLexpl}
	\begin{align}
	\mathcal{A}_L &= \frac{2^{\ell-1}}{\ell!}\hat{\partial}_L\!\left[\frac{1}{r}\left(\frac{t-r-s}{2}\right)^\ell\left(\frac{t-r-s}{2}-{r'}\right)^\ell\right]\,,\\
	\mathcal{B}_L &= \frac{2^{\ell-1}}{\ell!}\hat{\partial}_L\!\left[\frac{1}{r}\left(\frac{t+r-s}{2}\right)^\ell\left(\frac{t+r-s}{2}-{r'}\right)^\ell\right]\,.
	\end{align}
\end{subequations}
The fact that $\mathcal{A}_L$ and $\mathcal{B}_L$ are actually identical follows immediately from Eq.~(A36) in~\cite{BD86}. At last, we can commute the multi-spatial derivative operator $\hat{\partial}_L$ in $\mathcal{A}_L$ and $\mathcal{B}_L$ with the integration over ${r'}$ in~\eqref{eq:h3d2}. Indeed, the contributions coming from the differentiation of the bound $(t\mp r-s)/2$ are seen to be zero from the structure of $\mathcal{A}_L$ and $\mathcal{B}_L$ respectively. So, finally, posing
\begin{align}\label{eq:R}
R(\rho, s) = \rho^\ell \int_0^\rho \dd{r'} \frac{(\rho-{r'})^\ell}{\ell!}\left(\frac{2}{{r'}}\right)^{\ell-1}\!{r'}^B N({r'}, s+{r'})\,,
\end{align}
we get our final result in 3 dimensions,
\begin{align}\label{eq:h3dfinal}
&h_L = \int_{-\infty}^{t-r}\dd s \,\hat{\partial}_L\left[\frac{R\bigl(\frac{t-r-s}{2},s\bigr)-R\bigl(\frac{t+r-s}{2},s\bigr)}{r}\right]\,.
\end{align}
This is exactly the result given by Eq.~(6.4) in~\cite{BD86}.

\bibliography{ListeRef_DiffIR.bib}

\end{document}